\documentclass[aps,prd,reprint,superscriptaddress,nofootinbib]{revtex4-2}
\usepackage{amsmath,amsfonts,amssymb,bm}
\usepackage{graphicx}
\usepackage{bbold}
\usepackage[normalem]{ulem}
\usepackage[dvipsnames,usenames]{xcolor}
\usepackage[colorlinks,citecolor=blue]{hyperref}
\usepackage{orcidlink}

\begin{document}

\title{Neutrino quantum kinetics for fast flavor conversion in a time-dependent environment}

\author{Zewei Xiong\orcidlink{0000-0002-2385-6771}}
\email[Email: ]{z.xiong@gsi.de}
\affiliation{GSI Helmholtzzentrum {f\"ur} Schwerionenforschung, Planckstra{\ss}e 1, 64291 Darmstadt, Germany}

\author{Meng-Ru Wu\orcidlink{0000-0003-4960-8706}}
\email[Email: ]{mwu@as.edu.tw}
\affiliation{Institute of Physics, Academia Sinica, Taipei 11529, Taiwan}
\affiliation{Institute of Astronomy and Astrophysics, Academia Sinica, Taipei 10617, Taiwan}
\affiliation{Physics Division, National Center for Theoretical Sciences, Taipei 10617, Taiwan}

\date{\today}

\begin{abstract}
Fast flavor conversions (FFCs) of neutrinos, driven by the fast flavor instability (FFI), can reshape the neutrino flavor content in dense astrophysical environments such as core-collapse supernovae and neutron star mergers.
Most studies of FFCs adopt a two-step approach, in which a flavor-unstable state containing deep electron-minus-heavy-flavor lepton number (E-XLN) angular crossings is first constructed and subsequently evolved.
Because realistic crossings should instead develop gradually through neutrino transport, the validity of such setups has been called into question.
We investigate this issue by solving the neutrino quantum kinetic equations with self-consistent collisional rates in a spherically symmetric supernova background whose electron fraction evolves in time through a sequence of stages, starting from a configuration free of E-XLN crossings.
We find that the evolution proceeds through three characteristic episodes.
In the shallow-crossing episode, FFCs develop from marginally unstable, shallow crossings, carrying small-scale structures consistent with linear stability analysis.
In the near-crossing-elimination episode, the balance between collisions and FFCs keeps the system in a near-quasistationary state in which the emerging crossings are continuously eliminated, so that the strongly unstable regime is never reached.  
In the swapping episodes, the E-XLN reverses sign and a dynamically propagating flavor-swap E-XLN zero surface forms.
We find that the evolved flavor content at the end of different time stages broadly agrees with the quasistationary solutions obtained in the corresponding two-step models adopting fixed matter backgrounds. 
In addition, we investigate the robustness of the effective classical transport (ECT) framework that adopts subgrid flavor redistribution using different parametrized prescriptions. 
Notably, except during the swapping episodes, where FFCs are not driven by FFIs, the ECT models reproduce the QKE results with good quantitative accuracy, consistent with our previous study adopting the two-step setup [Z. Xiong \emph{et~al}., \href{https://doi.org/10.1103/PhysRevLett.134.051003}{Phys. Rev. Lett. \textbf{134}, 051003 (2025)}]. 
\end{abstract}

\maketitle
\graphicspath{{./}{figures/}}

\section{Introduction}
\label{sec:introduction}

Neutrinos play a central role in core-collapse supernovae (CCSNe) and neutron star mergers (NSMs), carrying away a large fraction of the gravitational binding energy released and influencing the energy transport.
Their interactions with matter and with each other can influence the dynamics, nucleosynthesis, and the associated observable signatures.
In particular, the phenomenon of neutrino flavor conversion, whereby neutrinos change their flavor states as they propagate, can significantly modify the energy spectra and angular distributions of the emitted neutrinos.
In addition to the vacuum and matter-induced flavor oscillations, the coherent neutrino–neutrino forward scattering can trigger fast flavor instability (FFI; see e.g., Refs.~\cite{volpe2024neutrinos,johns2025neutrino} for recent reviews), which leads to fast flavor conversions (FFCs).
These occur on timescales and length scales associated with the local neutrino–neutrino interaction strength, potentially orders of magnitude shorter than scales of conventional collective oscillations, neutrino interaction cross sections, and hydrodynamics, and can take place in a wide range of regions within these astrophysical environments 
\cite{sawyer2016neutrino,wu2017fast,wu2017imprints,abbar2019occurrence,delfanazari2019linear,nagakura2019fastpairwise,morinaga2020fast,abbar2020fast,glas2020fast,nagakura2021where,abbar2021characteristics,harada2022prospects,akaho2023collisional,mukhopadhyay2024time,nagakura2025neutrino,xiong2025occurrence,froustey2026neutrino,urquilla2026neutrino}.

The occurrence of FFCs is closely linked to the angular distributions of neutrinos and antineutrinos.
In particular, a sufficient and necessary condition for their onset is the presence of electron-minus-heavy-flavor lepton number (E-XLN) crossings, where the angular distributions of electron neutrinos and antineutrinos intersect  \cite{morinaga2022fast,dasgupta2022collective,fiorillo2025collective}.
Such crossings are expected to arise naturally in the decoupling regions, where complex transport effects and multidimensional hydrodynamic instabilities produce anisotropic neutrino emission.
Once triggered, FFCs can efficiently drive the neutrino gas toward flavor equilibration, modifying the lepton number transport.

The nonlinear and multidimensional nature of FFCs, as well as the associated scale separation, makes the quantification of their impact on astrophysical environments an open and active area of research \cite{xiong2020potential,fujimoto2022explosive,nagakura2023roles,ehring2023fast,ehring2023fast1, george2020fast,li2021neutrino,just2022fast,fernandez2022fast,nagakura2023global,nagakura2023connecting,nagakura2023basic,froustey2024neutrino,grohs2024twomoment, nagakura2024neutron,mori2025threedimensional,qiu2025neutrino,qiu2025impact,wang2025effect,wang2025effect2,lund2025angledependent,akaho2026bifurcated,gogilashvili2026neutrino,friedland2026solarsystem}.
Despite significant progress in recent years, understanding the interplay between neutrino flavor conversions and hydrodynamics remains one of the key challenges in developing a complete picture of neutrino behavior and their astrophysical implications.

Aiming to address the challenge of the scale separation, several recent studies have introduced subgrid neutrino transport models~\cite{nagakura2024bgk,xiong2025robust,akaho2025comparative} to emulate the outcomes of FFCs without solving the quantum kinetic equations (QKEs)  \cite{sigl1993general,vlasenko2014neutrino,volpe2015neutrino,blaschke2016neutrino}.
These models typically utilize prescriptions that parameterize the flavor equilibration process by enforcing local flavor mixing or adjusting neutrino lepton number fluxes, guided by local periodic-box QKE simulations \cite{martin2020dynamic,bhattacharyya2021fast,bhattacharyya2020latetime,wu2021collective,richers2021particleincell,richers2021neutrino,zaizen2021nonlinear,abbar2022suppression,richers2022code,bhattacharyya2022elaborating,grohs2023neutrino,zaizen2023simple,xiong2023evaluating,richers2024asymptoticstate,george2024evolution,goimilia2025steady}.
Adopting these subgrid models allows multidimensional CCSN and NSM simulations to approximately account for the influence of FFCs on neutrino transport and matter composition without directly solving the complete QKEs. 
Encouragingly, 
such implementations in state-of-the-art simulations have already demonstrated stable performance, marking an important step toward bridging microscopic flavor oscillation physics and macroscopic astrophysical modeling \cite{lund2025angledependent,akaho2026bifurcated}.

Meanwhile, several groups have also investigated FFCs by solving the QKEs under realistic CCSN background profiles spanning over a global length of tens of kilometers~\cite{nagakura2022timedependent,shalgar2023neutrino,shalgar2023neutrino1,nagakura2023connecting,nagakura2023basic,xiong2024fast}.
In contrast to the subgrid models, these global simulations either adopt quenched neutrino-neutrino interaction strength or take spatial resolution much lower than the FFC length-scale to mitigate the scale separation problem. 
In particular, Ref.~\cite{xiong2025robust} found excellent agreement between the solutions obtained from the global QKE solutions and those from the subgrid models, demonstrating the effectiveness of subgrid models.

Many local-box and global simulations involve a two-step approach:
The first step sets up the initial condition that is flavor unstable, i.e., containing E-XLN crossings, and the second step simply evolves the QKE without further involving any other time-dependent variation of the system.
The motivation of such a two-step approach is clear:
It allows the flavor-evolved system to settle into a quasi-stationary state.
For local periodic-box simulations, the quasi-stationary solutions are the basis of the aforementioned prescriptions.
For global simulations, this approach allows direct comparison of the outcomes of flavor-evolved and flavor-suppressed neutrino fields under identical physical conditions, making it possible to assess how FFCs change key quantities such as lepton number fluxes, neutrino spectra, or heating rates.

However, the set-up of the two-step approach has been called into question in a number of recent works~\cite{fiorillo2024fast,johns2025subgrid,johns2025localequilibrium,urquilla2025testing}.
The major concern is as follows.
In realistic astrophysical environments, the E-XLN angular crossings should develop gradually over time since the angular distributions of neutrinos and antineutrinos evolve continuously through neutrino transport in response to the hydrodynamical evolution.
Hence, the flavor instability should be weak when a shallow crossing first appears.
Should the resulting FFCs eliminate the shallow E-XLN crossing once a weak instability develops, the system would never have entered the strongly-unstable regime adopted in the two-step approach.

This argument points to several potential limitations of the two-step approach and the related questions in understanding FFCs. 
The first limitation arises from the possibility of artificial conversions caused by the sudden activation of flavor oscillations associated with strong FFIs.
This artifact drives unrealistically strong flavor mixing that may not accurately represent the gradual, self-consistent development of instabilities in a continuously evolving system. 
The response of the FFIs to the gradual changes of weak E-XLN crossings remains poorly understood. 

The second limitation is that the lack of slow variation in local conditions may prevent phenomena that arise when the system gradually transitions between different regimes. 
For instance, the flavor swap through the electron lepton zero surface \cite{zaizen2024fast} or the continued flavor evolution without further presence of E-XLN crossing \cite{fiorillo2024fast,liu2024quasisteady,urquilla2025testing} are found to emerge when the local conditions gradually transition from $\nu_e$-dominated to $\bar{\nu}_e$-dominated configurations in space or time. 
It remains to be seen whether these phenomena naturally arise in a set-up where the transition is driven by neutrino transport instead of artificially introduced by external source terms or
boundary conditions.

The third limitation concerns the validity of the prescriptions used in subgrid models.
As these prescriptions are formulated to represent the local flavor-equilibrated state based on the quasistationary solutions obtained from two-step local QKE simulations, their general applicability in systems where E-XLN crossings gradually appear should be further tested.

\begin{figure}[!hbt]
\centering
\includegraphics[width=\columnwidth]{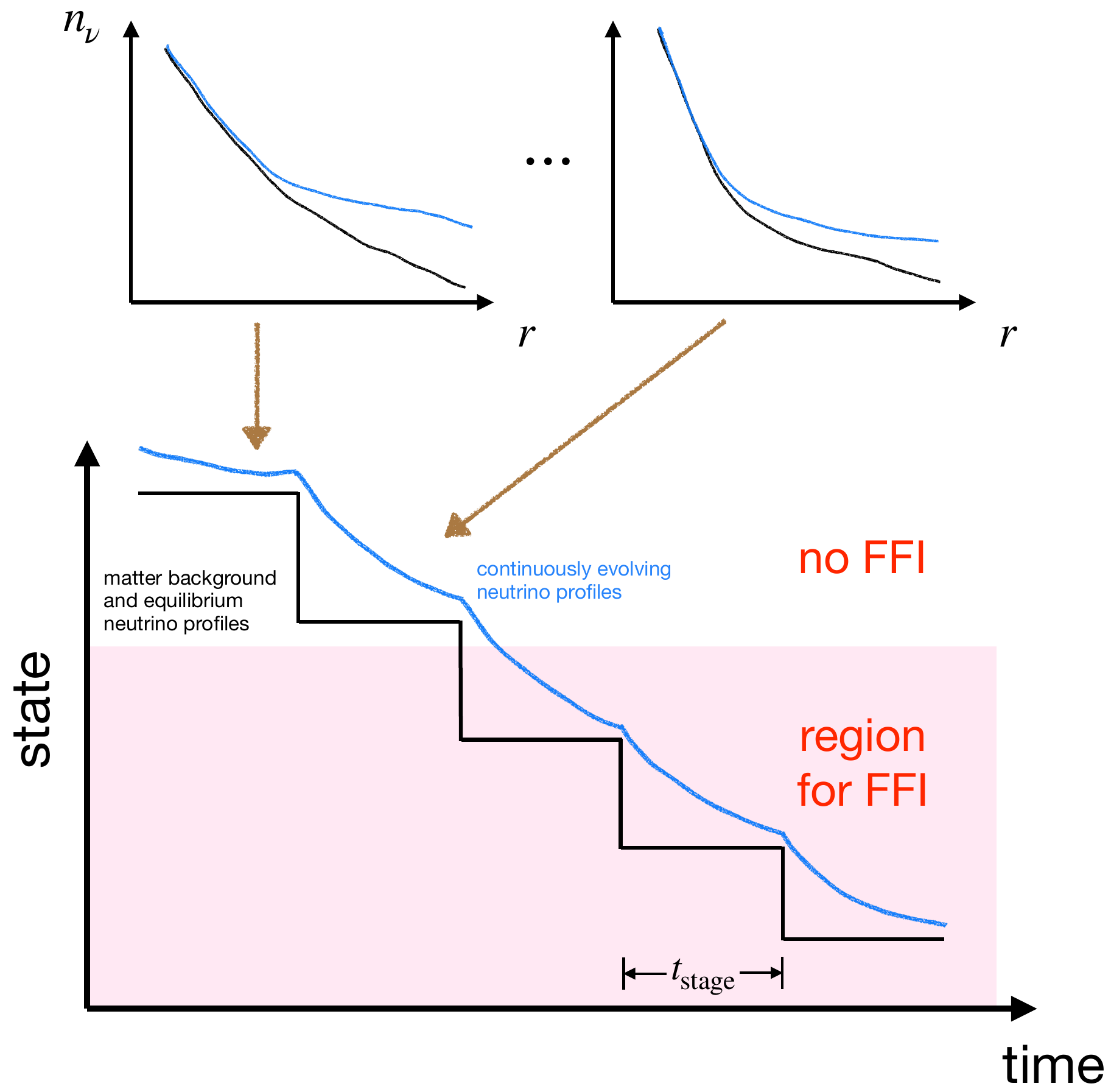}
\caption{\label{fig:diagram} Schematic illustration of the staged time-dependent background used to study the near-quasistationary evolution with shallow angular crossings. Starting from an initial configuration in which no E-XLN angular crossing is present in the neutrino distributions, the system evolves through a sequence of quasi-stationary stages, each characterized by a fixed matter background that persists for a finite duration $t_{\rm stage}$.}
\end{figure}

To explore these effects and answer the above questions, we extend our previous simulation based on stationary spherically symmetric supernova profiles to a setup including a time-dependent matter background, allowing the conditions for FFCs to evolve continuously.
The overall setup is illustrated schematically in Fig.~\ref{fig:diagram}.
We begin from an initial configuration in which no E-XLN angular crossing is present in the neutrino distributions.
The system is then evolved through a sequence of different stages, each characterized by a fixed matter background that is kept constant over a finite duration $t_{\rm stage}$.
Within each stage, the neutrino distributions are allowed to relax under collisional processes, enabling them to adjust self-consistently to the local conditions.
After $t_{\rm stage}$, the matter background is replaced by a new profile slightly modified from the previous one, to mimic the slow evolution of the hydrodynamic environment.

This stepwise change of matter background allows the neutrino field to coevolve smoothly, without introducing any FFIs in the initial condition.
As we will show later, when the matter and neutrino fields evolve, the E-XLN angular distributions gradually change, and shallow crossings naturally develop in certain regions.
The evolution of the neutrino angular distribution is then driven by the interplay of FFCs, collisions, and advection, leading to near-quasistationary evolution of the system, which remains marginally unstable.
Moreover, we will compare the result obtained at the end of each stage to the quasistationary solution of the corresponding two-step model, whose initial condition is flavor unstable as determined by transport without including flavor oscillations~\cite{xiong2024fast}.
The QKE solution will also be compared with the subgrid models adopting the effective classical transport (ECT) framework~\cite{xiong2025robust}.

The paper is organized in the following structure.
We provide the descriptions and parameters of the staged matter background models for the FFCs in Sec.~\ref{sec:models}.
The general features of the time-dependent FFC evolution are highlighted in Sec.~\ref{sec:global}, where we also classify the evolution into three characteristic episodes.
The detailed dynamics of each episode are analyzed in Sec.~\ref{sec:episodes}, supported by results from linear stability analysis (LSA) \cite{banerjee2011linearized,izaguirre2017fast,capozzi2017fast,capozzi2019fast,yi2019dispersion,fiorillo2024theory,fiorillo2024theory1} and the spectrogram analysis~\cite{xiong2024fast} presented in Sec.~\ref{sec:spectrogram}.
We discuss the impact of the attenuation method in Sec.~\ref{sec:attenuation}.
In Sec.~\ref{sec:comparison}, we compare the quasistationary solutions obtained from two-step models and the subgrid prescriptions against the time-dependent QKE solutions, and we conclude in Sec.~\ref{sec:conclusion}.
Natural units with $\hbar=c=k_B=1$ are adopted throughout the paper.

\section{Models}
\label{sec:models}

\subsection{Time-dependent matter background}
In this work, we take the same spherically symmetric CCSN background profile used in Ref.~\cite{xiong2024fast}, taken at 252~ms after the core-bounce of a CCSN model simulated with the \textsc{agile-boltztran} code~\cite{mezzacappa1993type,mezzacappa1993stellar,mezzacappa1993numerical,liebendorfer2001conservative,liebendorfer2004finite} for a progenitor star with zero-age main sequence mass of 25 $M_\odot$ and solar metallicity, as the baseline profile; see Ref.~\cite{xiong2024fast} for detailed descriptions of the model.

Based on this baseline profile, we introduce modifications to the electron fraction ($Y_e$) to represent variations of the neutron-to-proton ratio that may arise from multidimensional dynamics or convection, which can create conditions favorable for FFCs. 
Specifically, the original $Y_e$ profile is multiplied by a radially dependent factor
\begin{equation}
b_{Y_e}(t) = 1-b_1(t)/[1+\exp((r-b_2)/(1 ~\mathrm{km}))]. 
\end{equation}
Here $b_1(t)$ sets the size of attenuation, whereas the choice $b_2=42$~km confines the modification mainly to the region near the proto-neutron star.
In order to implement the staged matter background evolution discussed in Introduction, the value of $b_1$ is first gradually increased in steps of $0.01$ at the end of each stage, from $b_1=0.04$ to $b_1=0.14$, followed by a sequence of incremental decreases of $0.01$ after each stage, returning to $b_1=0.04$.
The variation of $b_1$ in this range allows the neutrino profiles to transition from the conditions dominated by $\nu_e$'s to those dominated by $\bar\nu_e$'s. 
An improvement over our previous work in \cite{xiong2024fast} is that the neutron and proton abundances are recalculated consistently with the modified $Y_e$ using the same DD2 nuclear equation of state~\cite{typel2010composition} adopted in the CCSN model.
This update increases (decreases) $\mu_p$ ($\mu_n$) from their original values.
Hence, the maximum of $b_1$ required to achieve a similar degree of $\bar\nu_e$ dominance considered in Model V of Ref.~\cite{xiong2024fast} is only 0.14, roughly half of the number $0.25$ used therein.
As a result, the conditions for triggering the FFI may more closely resemble what can be realized in realistic astrophysical settings.
The default duration of each stage, $t_{\rm stage}$, is set to 0.16~ms. 
Two additional cases with $t_{\rm stage}=0.08$ and $0.32$~ms are also considered. 
The value of $t_{\rm stage}$ and other parameters (see next Section) used in different models are listed in Table~\ref{tab:parameters}.

\subsection{Neutrino quantum kinetics}
Taking the two-flavor scheme, the evolution of the neutrino and antineutrino flavor-density matrices, $\varrho$ and $\bar\varrho$, is governed by
\begin{align}
    ( \partial_t + v_r \partial_r + \frac{1-v_r^2}{r}\partial_{v_r}) \varrho & = -i[\mathbf H, \varrho] +\mathbf C,
    \label{eq:eom_nu}
\end{align}
and
\begin{align}
    ( \partial_t + v_r \partial_r +\frac{1-v_r^2}{r}\partial_{v_r}) \bar\varrho & = -i[\bar{\mathbf H}, \bar\varrho] +\bar{\mathbf C},
    \label{eq:eom_nubar}
\end{align}
with
\begin{equation}
    \varrho = \begin{bmatrix} \varrho_{ee} & \varrho_{ex} \\ \varrho_{ex}^* & \varrho_{xx} \end{bmatrix} \text{~and~}
    \bar\varrho = \begin{bmatrix} \bar\varrho_{ee} & \bar\varrho_{ex} \\ \bar\varrho_{ex}^* & \bar\varrho_{xx} \end{bmatrix}.
\end{equation}
The normalization is chosen such that $n_{\nu_i}(r,t)=\int dE\,dv_r\,\varrho_{ii}$, with $i=e,x$ and $x$ denoting a heavy-lepton flavor.

In this work, the coherent propagation Hamiltonian $\mathbf H$ that we consider contains the vacuum mixing term 
\begin{equation}
    \mathbf H_{\rm vac}(E) =
    \frac{\delta m^2}{4E}
    \begin{bmatrix}
    -\cos 2\theta_V & \sin 2 \theta_V \\ \sin 2 \theta_V & \cos 2 \theta_V
    \end{bmatrix}
\end{equation}
with the mass-squared difference $\delta m^2$ and the mixing angle $\theta_V$, and the neutrino self-induced term
\begin{equation}
    \mathbf H_{\nu\nu}(v_r) = \sqrt{2} G_F \int d E'\, d v_r' (1-v_r v_r') (\varrho-\bar\varrho^*),
\end{equation}
which accounts for coherent forward scattering among neutrinos.
Although the matter term may affect the behavior of FFCs \cite{sigl2022simulations,bhattacharyya2025role,zaizen2026fast,fiorillo2026flavomons}, we do not include it explicitly but use a suppressed effective mixing angle to partially account for its effect.
We set $\theta_V=10^{-6}$ and $\delta m^2=8\times 10^{-5}~\mathrm{eV}^2$ as in Ref.~\cite{xiong2024fast}.
For the collisional term $\mathbf C$, we also adopt the same treatment as in Ref.~\cite{xiong2024fast} to include the charged-current neutrino emission and absorption (EA) on nucleons, isoenergetic neutrino nucleon scattering (NNS), inelastic neutrino scattering on electrons and positrons (NES), and neutrino pair reactions (PR), as listed in Table~I therein.

\subsection{Numerical setup}
We solve the above QKEs with \textsc{cose$\nu$}, using the numerical framework of Ref.~\cite{xiong2024fast}. 
The simulation domain between the inner boundary $r_{\rm ib}=20$~km and the outer boundary $r_{\rm ob}=80$~km is uniformly discretized by $N_r$ radial grids. 
At $r_{\rm ib}$, the phase-space distributions of neutrinos of all species are determined by the Fermi-Dirac distribution functions whose chemical potentials are given by weak equilibrium for the included NES and PR reactions. 
At $r_{\rm ob}$, located near the CCSN shock, 
the free-streaming boundary condition is adopted for forward propagating neutrinos with $v_r\geq 0$, while no incoming neutrino population is imposed at the outer boundary. 
We take $N_E=15$ discretized logarithmically uniform energy grids between 1~MeV and 100~MeV and $N_{v_r}=100$ linearly uniform angular grids between $v_r=-1$ and 1 for both $\varrho$ and $\bar\varrho$.

For the neutrino self-induced term $\mathbf H_{\nu\nu}$, we reduce its strength by multiplying it with an artificial attenuation function
\begin{equation}
a_{\nu\nu}(r) = a_1/[1+\exp((30~\rm{km}-r)/(2.5~\rm{km}))]   
\end{equation}
as in Ref.~\cite{xiong2024fast} to alleviate the computational demand required for fully resolving the FFC length scale. 
With this choice, $a_{\nu\nu}(r)\approx a_1$ for $r>35$~km and $a_{\nu\nu}(r)> 0.1a_1$ at $r>25$~km.
We take three different choices of $a_1$, $N_r$, and the fixed time step size $\Delta t$ to investigate the dependence of the result on the attenuation parameter:   
$(a_1, N_r, \Delta t)=(4\times 10^{-3}, 50000, 1.6~{\rm ns})$, $(2\times 10^{-3}, 50000, 1.6~{\rm ns})$, and $(10^{-3}, 25000, 3.2~{\rm ns})$.  
The corresponding models are named QKE-H, QKE-M, and QKE-L, respectively. 
Flavor mixing is seeded only by the nonzero vacuum mixing; no additional perturbation is applied to the initial state.

Table~\ref{tab:parameters} lists the parameters used in all considered models.
These fiducial models always include the off-diagonal elements for the collision terms. 
As will be discussed later, we restart a few models at intermediate stages to investigate the effects of decoherence on the FFCs. 
For those models, the off-diagonal elements of $\mathbf C$ are manually set to zero after being restarted.

\begingroup
\begin{table}[t]
\begin{ruledtabular}
    \centering
    \caption{\label{tab:parameters} Varied parameters used in models investigated in this work. 
    For all models, we fix $r_{\rm ib}=20$~km,  $r_{\rm ob}=80$~km, $b_2=42$~km, $\delta m^2=8\times 10^{-5}~{\rm eV}^2$, $\theta_V=10^{-6}$, $N_{v_r}=100$, and $N_E=15$. 
    The values of $b_1$ are changed from one stage to the next stage, increasing from 0.04 to 0.14 with the increment of 0.01 and then decreasing to 0.04 with the decrement of 0.01 (see text for details).}
    \begin{tabular}{lcccc}
        model & $a_1$ & $t_{\rm stage}$~[ms] & $N_r$ & $\Delta t$~[ns]~\\\hline
        QKE-L       & $10^{-3}$         & 0.16 & 25000 & 3.2 \\
        QKE-M       & $2\times 10^{-3}$ & 0.16 & 50000 & 1.6 \\
        QKE-H       & $4\times 10^{-3}$ & 0.16 & 50000 & 1.6 \\
        QKE-L-long  & $10^{-3}$         & 0.32 & 25000 & 3.2 \\
        QKE-H-short & $4\times 10^{-3}$ & 0.08 & 50000 & 1.6 \\\hline
        model & $\tau$ & $t_{\rm stage}$~[ms] & $N_r$ & $\Delta t$~[ns]~\\\hline
        ECT         & Eq.~\eqref{eq:relaxation_tau} & 0.16 & 250 & 32 \\
        ECT-a0.1    & Eq.~\eqref{eq:relaxation_tau1}& 0.16 & 250 & 32 \\
        ECT-box     & Eq.~\eqref{eq:relaxation_tau} & 0.16 & 250 & 32 \\
    \end{tabular}
\end{ruledtabular}
\end{table}
\endgroup

\subsection{Illustrative quantities}
\label{sec:illus_quant}
For clarity in presenting our models, we define several quantities as follows. 
All quantities are time-dependent so we suppress their explicit dependence on $t$ for simplicity.
The energy-integrated density matrices are defined by
\begin{equation}\label{eq:avgrho}
    \langle\varrho\rangle_E(r,v_r) = \int d E\, \varrho(r,E,v_r).
\end{equation}
The $\nu_e$ number fraction representing the extent of flavor equilibration is defined as
\begin{equation}\label{eq:chi}
    \chi_{\nu_e}(r,v_r)
    = \frac{\langle\varrho_{ee}\rangle_E(r,v_r)}{\langle\varrho_{ee}\rangle_E(r,v_r)+\langle\varrho_{xx}\rangle_E(r,v_r)}.
\end{equation}
The phase (angle in the complex plane) of the off-diagonal element is defined as
\begin{equation}
    \phi_{ex}(r,v_r) = \arg[\langle \varrho_{ex} \rangle_E \, {\rm sgn}(\langle \varrho_{ee} \rangle_E-\langle \varrho_{xx} \rangle_E) ],
		\label{eq:phi_emu}
\end{equation}
where $\mathrm{sgn}$ denotes the sign function.

The E-XLN angular distribution is given by
\begin{equation}
    G(r,v_r)= \langle \varrho_{ee} \rangle_E-\langle \bar\varrho_{ee} \rangle_E-\langle \varrho_{xx} \rangle_E+\langle \bar\varrho_{xx} \rangle_E. 
\end{equation}
In addition, we define the coarse-grained E-XLN distribution
\begin{equation}
\tilde G(r,v_r)= \frac{1}{\Delta r}\int_{r-\Delta r/2}^{r+\Delta r/2} dr' G(r',v_r), 
\end{equation}
with $\Delta r=1.2$~km throughout the paper.
The coarse-grained $n$-th angular moment of the E-XLN distribution is
\begin{equation}
    I_n (r) = \int_{-1}^1 dv_r\, v_r^n \tilde G(r,v_r).
\end{equation}
For the E-XLN angular distribution, we define two integrals
\begin{align}
    I_\pm (r) & =\int_{-1}^1dv_r\, \tilde G (r,v_r) \Theta(\pm \tilde G(r,v_r)),
\end{align}
where $\Theta$ denotes the Heaviside step function.
Additionally, we define 
\begin{align}
    I_< = 
\left\{
  \begin{array}{rr}
    I_-, & {\rm if~}|I_-|<|I_+|; \\
    I_+, & {\rm otherwise};
  \end{array}
\right. \nonumber \\
    I_> = 
\left\{
  \begin{array}{rr}
    I_+, & {\rm if~}|I_-|<|I_+|; \\
    I_-, & {\rm otherwise}.
  \end{array}
\right.
    \label{eq:I_lessgtr}
\end{align}
Those two quantities allow us to evaluate the ``depth ratio'' of the E-XLN angular crossing based on the ratio of $|I_</I_>|$.
Another useful quantity to indicate the depth of the angular crossing, especially when flavor conversions have occurred, is
\begin{equation}
    \eta_{\rm osc}(r) =
\left\{
  \begin{array}{rr}
    \min_{v_r}(|\tilde G|), & {\rm if~no~crossing}; \\
    -|\min_{v_r}[{\rm sign}(I_0) \tilde G]|, & {\rm otherwise}.
  \end{array}
\right.
    \label{eq:lambda_osc}
\end{equation}
This quantity measures how flat or how shallow the angular crossing is if one exists.
If there is no crossing, it represents the distance of the angular distribution from zero.

\subsection{Effective classical transport models}
In addition to the QKE models, we also perform three simulations with the effective classical transport (ECT) framework outlined in \cite{xiong2025robust}. 
In short, the ECT models solve Eqs.~\eqref{eq:eom_nu} and \eqref{eq:eom_nubar} by excluding the commutators on the right hand sides, and use prescriptions from Refs.~\cite{zaizen2023simple,xiong2023evaluating}, which are formulated from the coarse-grained outcome obtained in local periodic-box QKE simulations, to redistribute neutrino distributions at any radius where an E-XLN angular crossing has developed. 

The intrinsic coarse-grained nature of the ECT scheme allows us to take a larger radial grid size no longer limited by the size of $\mathbf{H_{\nu\nu}}$ to drastically speed up the simulations.
We use 250 radial grids in the same simulation domain and a time step $\Delta t=32$~ns. 
Without the inclusion of the commutators, only the diagonal elements $\varrho_{ii}$ and $\bar\varrho_{ii}$ are evolved.
Correspondingly, the off-diagonal entries in the collision term are set to zero. 
All other settings in the ECT simulations are identical to those in the corresponding QKE calculations.

For flavor redistribution, at any radial grid points where sizable E-XLN crossings with $|I_</I_>| > 10^{-3}$ have developed, we replace $\varrho(r,v_r)$ and $\bar\varrho(r,v_r)$ for all energy grids by  
\begin{align}
    \varrho_{ii} & \rightarrow \varrho_{ii}+(\varrho_{ii}^{\rm (f)}-\varrho_{ii})\,{\rm min}(\Delta t/\tau, 1), \nonumber \\
    \bar\varrho_{ii} & \rightarrow \bar\varrho_{ii}+(\bar\varrho_{ii}^{\rm (f)}-\bar\varrho_{ii})\,{\rm min}(\Delta t/\tau, 1),
\end{align}
where $\varrho_{ii}^{\rm (f)}$ and $\bar\varrho_{ii}^{\rm (f)}$ are the equilibrated flavor content given by
\begin{align}\label{eq:ECT_power_update}
\varrho_{ee}^{\rm (f)}=\varrho_{ee}P+\varrho_{xx}(1-P), \nonumber \\
\varrho_{xx}^{\rm (f)}=\varrho_{ee}(1-P)+\varrho_{xx}P, 
\end{align}
with $P$ the energy-independent survival probability for (anti-)neutrinos given by the chosen prescription that eliminates the local E-XLN crossing (see below).
The parameter $\tau$ denotes the characteristic FFC relaxation timescale, estimated by 
\begin{equation}
    \tau=G_F^{-1}|I_+ I_-|^{-1/2}. 
\label{eq:relaxation_tau}
\end{equation}
When $\tau<\Delta t$, it reduces to the instantaneous redistribution scheme explored in Ref.~\cite{xiong2025robust}. 
When $\tau>\Delta t$, it effectively takes into account the finite relaxation effect that is considered in the Bhatnagar–Gross–Krook (BGK) subgrid approach~\cite{nagakura2024bgk}.
Note that in ECT models, quantities such as $I_>$, $I_<$, $I_+$, $I_-$, and $\eta_{\rm osc}$ are all defined based on grid values of $G$ without any further coarse-graining, since the ECT models already use coarse grids and flavor waves are absent by default. 

In addition to the fiducial ECT model, we perform another simulation using a much longer relaxation timescale  
\begin{equation}
    \tau'=10^4~G_F^{-1}|I_+ I_-|^{-1/2} .
    \label{eq:relaxation_tau1}
\end{equation}
This model is used to test the dependence of the ECT model on the relaxation timescale parameter. 

For the prescription that eliminates the E-XLN crossings used in Eq.~\eqref{eq:ECT_power_update}, we adopt two different prescriptions as in \cite{xiong2025robust} suitable for cases with single crossings.
The fiducial ECT model takes the power-1/2 prescription derived in \cite{xiong2023evaluating}.
For the two-flavor scheme, the energy-independent survival probability for (anti-)neutrinos is given by
\begin{equation}\label{eq:p_sur}
    P(v_r) = 
    \begin{cases}
        \frac{1}{2} & {\rm for~} v_<, \\
        1-\frac{1}{2}h(|v_r-v_c|/w) & {\rm for~} v_>,
    \end{cases}
\end{equation}
where the function $h(x)=(x^2+1)^{-1/2}$ and $v_c$ is the crossing velocity at which $G(v_c)=0$. 
In Eq.~\eqref{eq:p_sur}, $v_<$ ($v_>$) is defined as the $v_r$ range over which the absolute value of the positive or negative E-XLN, $|I_+|$ or $|I_-|$, is smaller (larger).
The parameter $w$ can be calculated to ensure the E-XLN conservation. 
The other scheme used in the ECT-box model is the ``box'' prescription \cite{zaizen2023simple}, which sets
\begin{equation}\label{eq:p_sur_box}
    P(v_r)=1-|I_</(2I_>)|,
\end{equation}
for $v_>$ while keeping $P(v_r)=1/2$ for $v_<$.

We note that the above procedure only strictly applies to cases with a single E-XLN crossing.
However, sometimes multiple shallow crossings may appear in ECT models.
When this occurs, we calculate the integral of $G$ over $v_r$ from -1 and 1 to the zero crossings closest to them as $I_{\rm left}$ and $I_{\rm right}$. 
If $I_{\rm left}$ and $I_{\rm right}$ have different signs, we associate $|I_{\rm left}|$ and $|I_{\rm right}|$ with $I_>$ and $I_<$ depending on their size, and use the same $P(v_r)$ from above for redistribution. 
For the region between those two zero crossings, we apply $P(v_r)=1/2$ as for the small side.  
On the other hand, if $I_{\rm left}$ and $I_{\rm right}$ have the same sign, $P(v_r)=1/2$ is applied only to the region between the two crossings, while the formulas associated with $v_>$ above are applied to the angular ranges covered by $I_{\rm left}$ and $I_{\rm right}$ based similarly on lepton-number conservation.

\begin{figure}
    \centering
    \includegraphics[width=0.98\columnwidth]{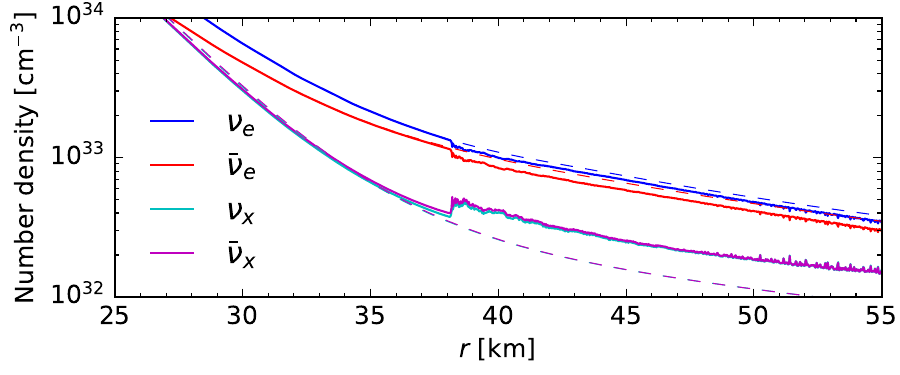}
\llap{\parbox[b]{1.2in}{\small (a) $t=0.32$~ms\\\rule{0ex}{1.1in}}} \includegraphics[width=0.98\columnwidth]{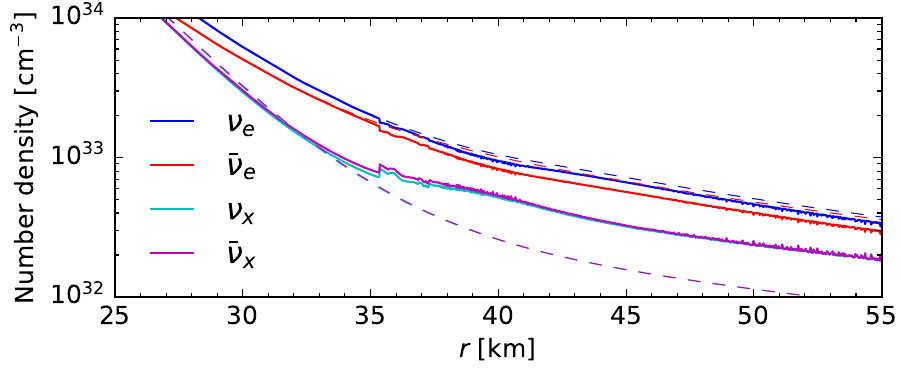}
\llap{\parbox[b]{1.2in}{\small (b) $t=0.64$~ms\\\rule{0ex}{1.1in}}}  \includegraphics[width=0.98\columnwidth]{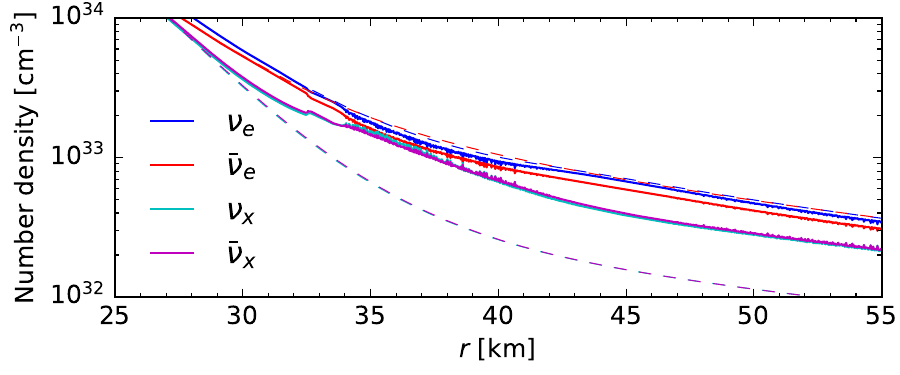}
\llap{\parbox[b]{1.2in}{\small (c) $t=0.96$~ms\\\rule{0ex}{1.1in}}}	 \includegraphics[width=0.98\columnwidth]{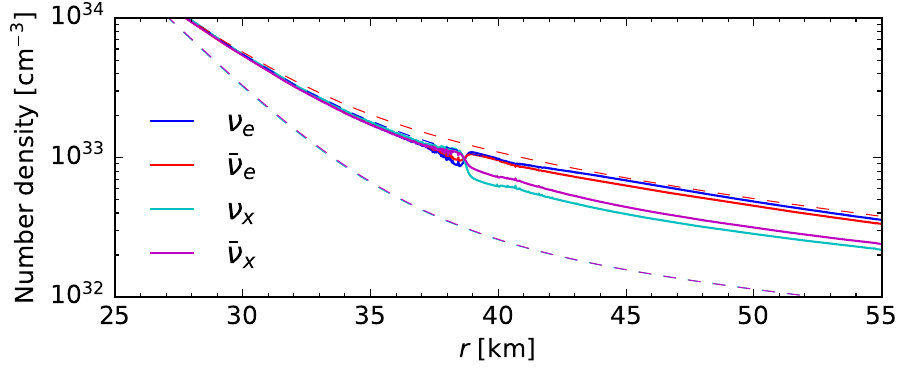}
\llap{\parbox[b]{1.2in}{\small (d) $t=1.28$~ms\\\rule{0ex}{1.1in}}}	  \includegraphics[width=0.98\columnwidth]{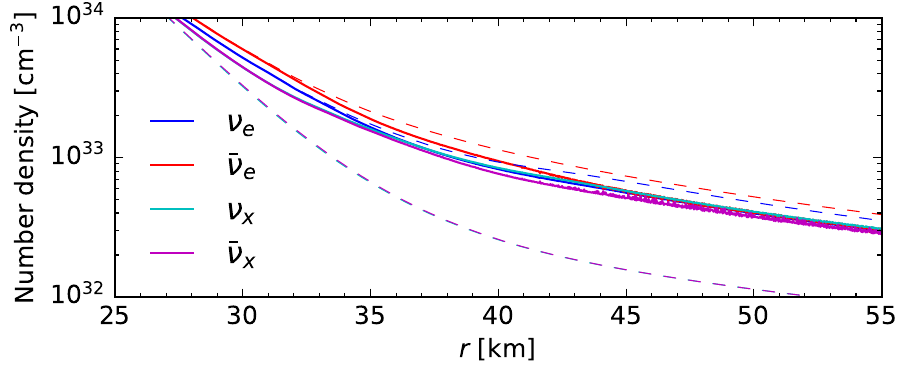}
\llap{\parbox[b]{1.2in}{\small (e) $t=1.6$~ms\\\rule{0ex}{1.1in}}}
    \caption{Radial profiles of neutrino number densities between $r=25$ and 55 km at selected times in model QKE-H.}
    \label{fig:rn_overall}
\end{figure}

\section{Global time-dependent evolution}
\label{sec:global}
In this section, we present the overall features of the simulation results and the qualitative trend of the evolution of the flavor-converted neutrino gases.
The quantities examined include the time-dependently evolved flavor content and the characteristics of the angular distributions at representative radii, as well as the radial profiles for the extent of flavor conversion.
We will also identify several important evolution episodes, but the details of the evolution will be analyzed in Sec.~\ref{sec:episodes}.
Our main discussion will focus on the fiducial QKE-H model. 
The curves of Models QKE-M and QKE-L are included in the figures to demonstrate that the results are insensitive to the attenuation factor $a_1$;
a detailed account of the attenuation effects is deferred to Sec.~\ref{sec:attenuation}.
The impact of the extended Models ``QKE-L-long'' and ``QKE-H-short'' with different $t_{\rm stage}$ will be addressed separately in Sec.~\ref{subsec:quasistationary}.

\subsection{Overall evolution of neutrino density profiles}
We first show in Fig.~\ref{fig:rn_overall} the number density profiles of different neutrino species taken at 5 snapshots for the model without oscillations (dashed lines) and the QKE-H model (solid lines) between 0.32 and 1.6~ms. 
When oscillations are turned off, the system transitions from $n_{\nu_e}>n_{\bar\nu_e}$ to $n_{\nu_e}<n_{\bar\nu_e}$ at $t\simeq 0.96$~ms as the parameter $b_1$ increases. 
For $n_{\nu_x}$ and $n_{\bar\nu_x}$, they are unaffected by the change of $b_1$ since they are produced in neutral-current processes, which are insensitive to the background $Y_e$ values.

When oscillations are turned on, these panels show that the pair-wise FFC occurs at outer radii first around the $\nu_e$ sphere at 40~km, 
leading to nearly equal amounts of flavor conversions between electron flavors and heavy-lepton flavors.
As time evolves, E-XLN crossings start to appear in regions inside the $\nu_e$ and $\bar\nu_e$ spheres. 
Consequently, the region affected by FFC extends to inner radii, resulting in increasingly higher $n_{\nu_x}$ and $n_{\bar\nu_x}$.
Remarkably, after the transition at $\sim 1$~ms, an interesting ``flavor swap'' feature, characterized by sharp changes of neutrino number densities at a specific radius, appears, illustrated by the profiles at $t=1.28$~ms. 
After this transient phase, FFCs operate continuously, giving rise to comparable number densities for different species at $t=1.6$~ms when $b_1$ reaches its maximal value of 0.14.

While the evolution of the density profiles at the second phase (when $b_1$ is sequentially decreased back to 0.04) qualitatively follows the inverse trend shown in Fig.~\ref{fig:rn_overall} and is hence not shown, there exists time-reversal asymmetry particularly around the times when the flavor swap feature appears, which will be discussed in detail in Sec.~\ref{subsec:swapping}.

\subsection{Radial dependence of the flavor conversion impact}
\label{subsec:radial}
\begin{figure*}[!hbt]
\centering
\includegraphics[width=0.45\textwidth]{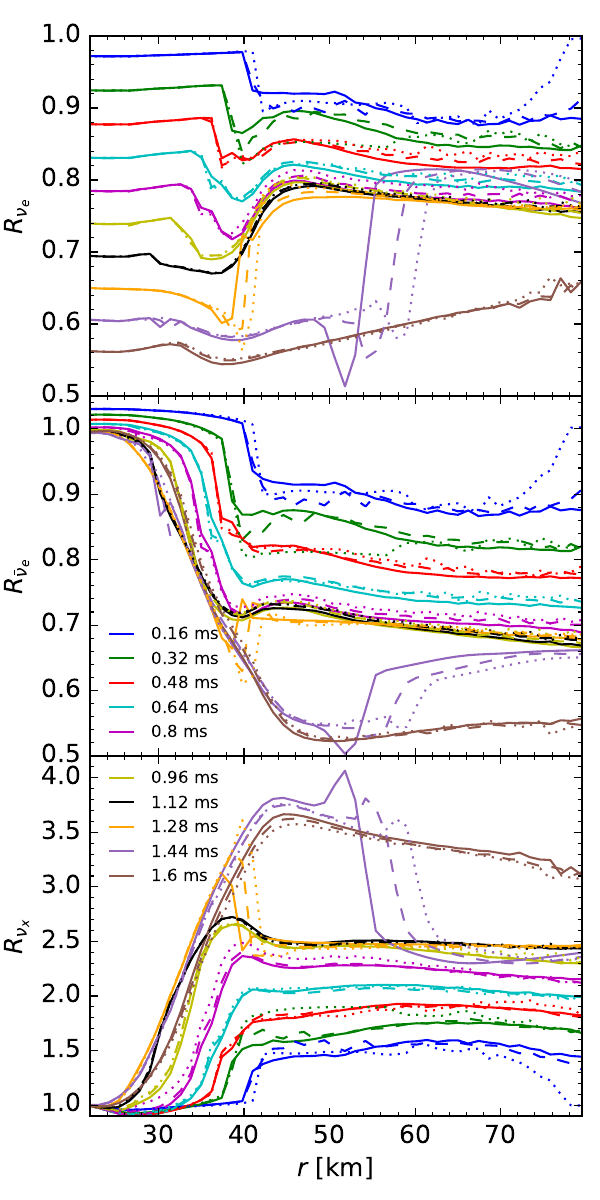}
\includegraphics[width=0.45\textwidth]{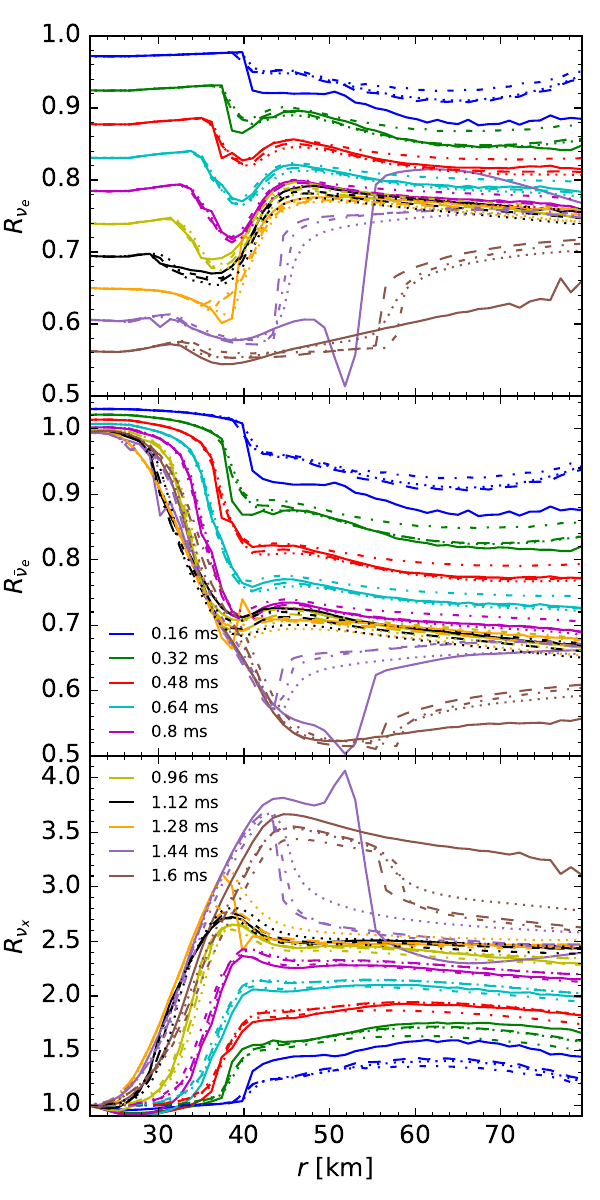}
\caption{\label{fig:rp__a421} Radial profiles for the ratio of neutrino number density over that at $t=0$.
Colors are for different simulation times $t$.
The solid, dashed, and dotted curves in the left panels are for models ``QKE-H'', ``QKE-M'', and ``QKE-L''.
The solid, dashed, dotted, and dashed-dotted-mixed curves in the right panels are for models ``QKE-H'', ``ECT'', ``ECT-a0.1'', ``ECT-box'', respectively.
For better illustration, the ratio $R_{\nu_e}$ ($R_{\bar{\nu}_e}$) at each time stage is plotted with an artificial offset of $-0.02$ ($-0.04$) relative to the previous time stage.
Without this offset, the ratios for $\nu_e$ and $\bar\nu_e$ at 20~km and $1.6$~ms are 0.74 and 1.39, respectively.
This offset helps avoid significant overlap among the curves in the plot.}
\end{figure*}

Let us now examine the impact of FFC on different neutrino species over the entire radial range.
To see this, we show in Fig.~\ref{fig:rp__a421} the number density ratio between simulation time $t$ and initial time $t=0$ as
\begin{equation}
    R_{\nu_i}(r,t) = \frac{n_{\nu_i}(r,t)}{n_{\nu_i}(r,t=0)},
\end{equation}
where $\nu_i=\nu_e,\,\bar\nu_e,\,\nu_x$, for 10 snapshots at the end of the first 10 stages.
The left panels compare the QKE-H (solid), QKE-M (dashed), and QKE-L (dotted) models, while the right panels compare the QKE-H model to the ECT models, which will be further discussed in Sec.~\ref{sec:comparison}.
Note that $R_{\nu_e}(r,t)$ and $R_{\bar\nu_e}(r,t)$ at different $t$ shown in the upper and middle panels are shifted by $-0.02$ and $-0.04$ relative to the previous stage, respectively, for better visibility.

In general, FFCs lead to enhancement in $n_{\nu_x}$ and reduction in $n_{\bar\nu_e}$ across regions where they have occurred.
Interestingly, the impact on $\nu_e$ can be two-fold, depending on the evolution stage.
At earlier times, e.g., $t<0.16$~ms, FFCs convert both electron flavors to the heavy-lepton ones outside the neutrinosphere, so $n_{\nu_e}$ simply decreases above the onset radii of FFIs similar to $n_{\bar\nu_e}$.
However, as the region inside the neutrinosphere starts to develop E-XLN crossings, FFCs work together with the collisional feedback effect~\cite{xiong2024fast}, which repopulates $\nu_e$ more efficiently than $\bar\nu_e$ around the neutrinosphere.
This effect leads to increasingly enhanced values of $R_{\nu_e}(r,t)$ over time at regions outside the onset radius of FFI.

When flavor swap appears, represented by $t=1.28$~ms and 1.44~ms, sharp transitions in $R_{\nu_e}$, $R_{\bar\nu_e}$, and $R_{\nu_x}$ appear.
Although here we only show two snapshots, this swapping interface is in fact generated at inner radii and propagates outward dynamically.
The left panels also indicate that the propagation of this interface depends on the adopted attenuation strength, which we discuss in Sec.~\ref{sec:attenuation}.

\subsection{Evolution of key quantities at representative radii and different episodes}
\label{subsec:representative}

\begin{figure*}[!hbt]
\centering
\includegraphics[width=0.98\textwidth]{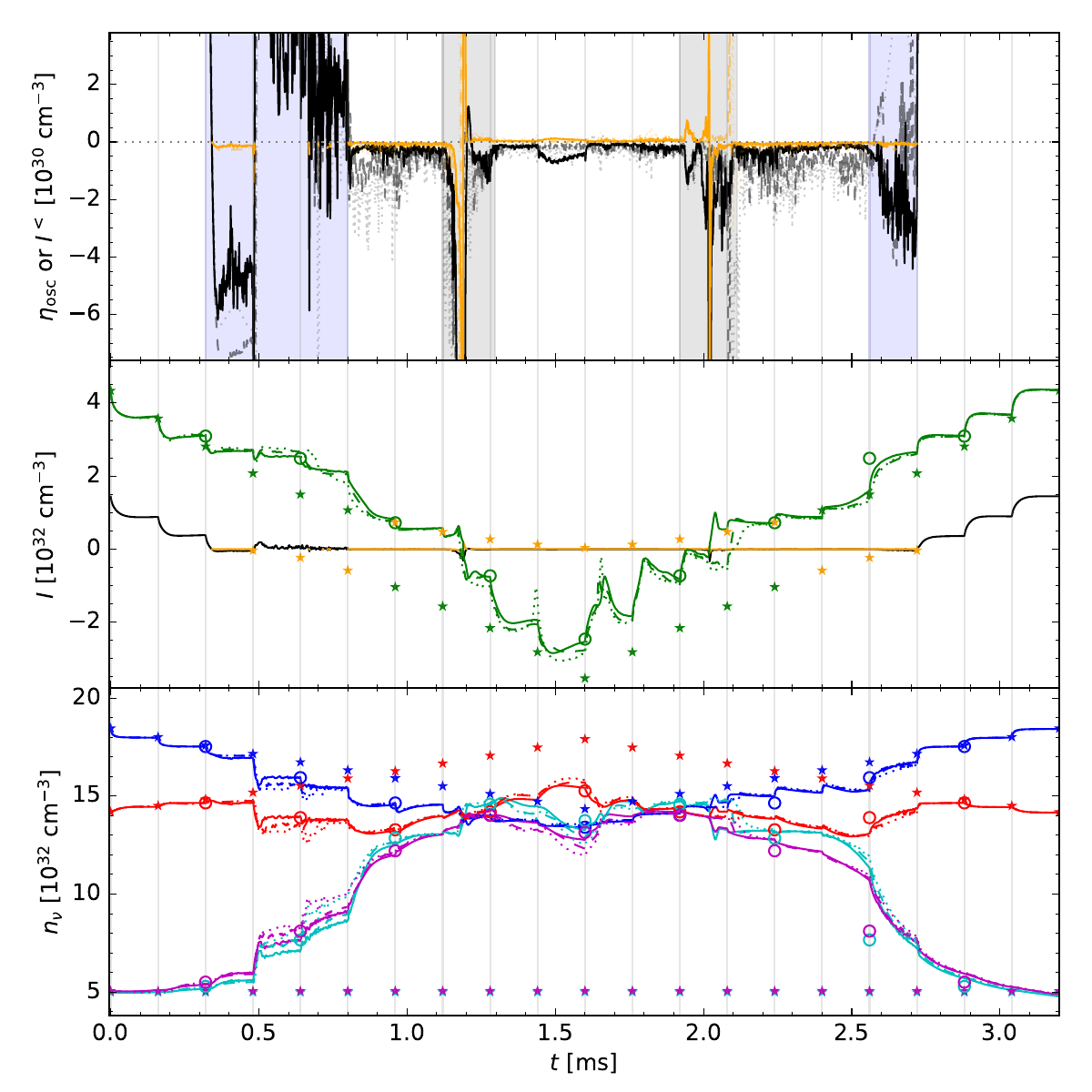}
\llap{\parbox[b]{6.2in}{\raggedright \small $b_1:$\\\rule{0ex}{4.45in}}}
\llap{\parbox[b]{6.25in}{\raggedright \small 4\%~~~5\%~~~6\%~~~7\%~~~8\%~~~9\%~~10\%~11\%~~12\%~~13\%~14\%~~13\%~~12\%~11\%~~10\%~~~9\%~~~8\%~~~7\%~~~6\%~~~5\%\\\rule{0ex}{4.3in}}}
\llap{\parbox[b]{13.2in}{\small (a)\\\vspace{1.9in}(b)\\\vspace{1.95in}(c)\\\rule{0ex}{2.4in}}}
\caption{\label{fig:lams_iz13_nopre}
Characteristics for neutrino angular distributions (a--b) and number densities of all species (c) at 36~km as functions of evolution time.
In (a)~and~(b), black, green, and orange curves are for $\eta_{\rm osc}$, $I_>$, and $I_<$, respectively.
The scale of $y$-axis in panel (a) is smaller than that in panel (b) to zoom in the details of $\eta_{\rm osc}$ and $I_<$.
The $I_<$ curves only exist when the coarse-grained E-XLN angular crossings appear.
In (a), the transparent blue and black bands indicate the shallow-crossing and swapping episodes, respectively.
In (c), blue, red, cyan, and magenta curves are for $n_{\nu_e}$, $n_{\bar\nu_e}$, $n_{\nu_x}$, $n_{\bar\nu_x}$, respectively.
In all panels, the gray vertical lines mark the time points at which the matter background changes to the next configuration.
The values of the $Y_e$ attenuation factor that governs the neutron-richness of the background matter, $b_1$, for each stage are labeled at the top of panel (b).
Solid, dashed, and dotted curves are for the ``QKE-H'', ``QKE-M'', and ``QKE-L'' simulations, respectively.
The solid stars are the same quantities calculated based on the asymptotic neutrino profiles obtained in simulations using static matter background indicated by the gray lines without including neutrino oscillations. 
The hollow circles are the asymptotic solutions of the corresponding two-step QKE-H model.}
\end{figure*}

\begin{figure*}[!hbt]
\centering
\includegraphics[width=0.98\textwidth]{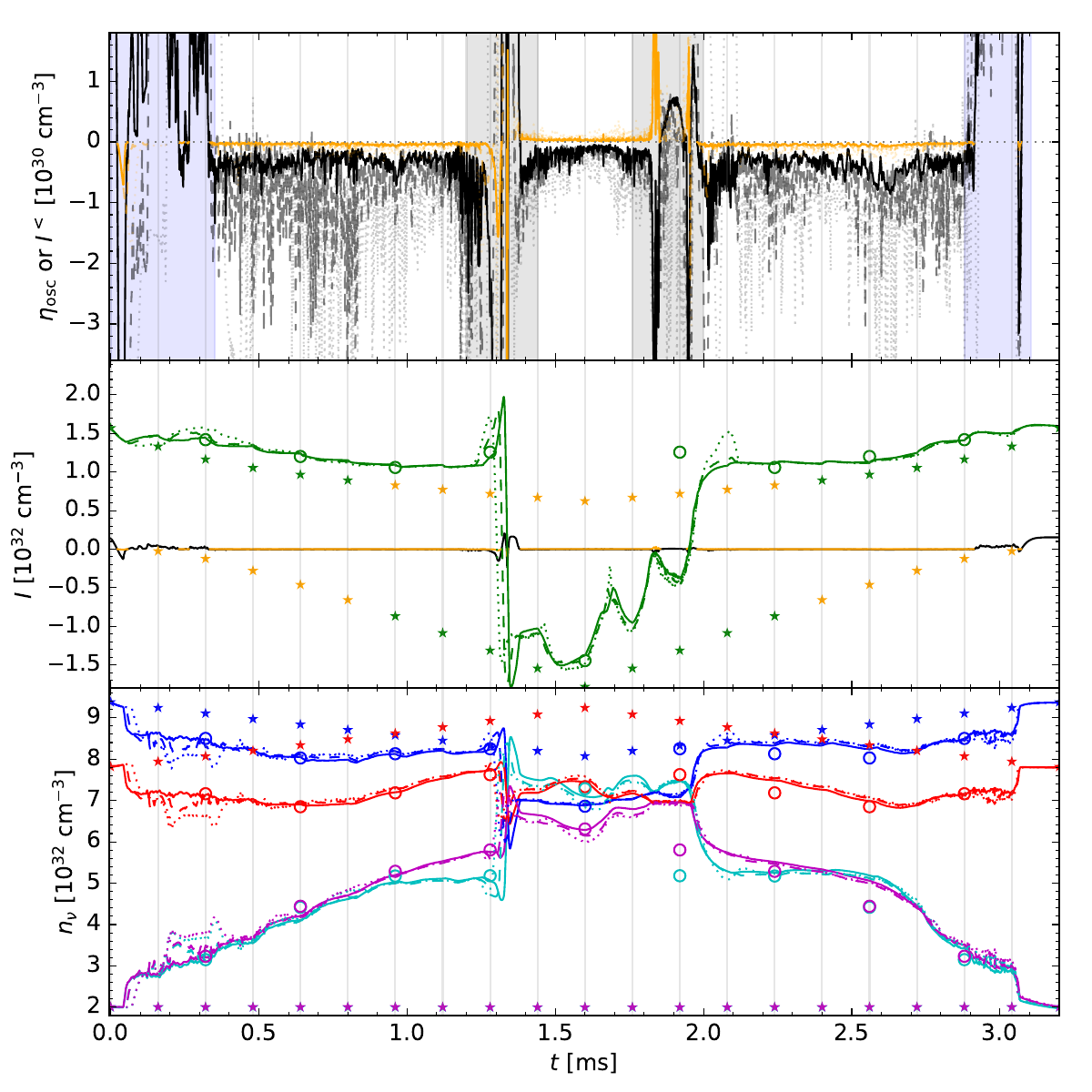}
\llap{\parbox[b]{6.2in}{\raggedright \small $b_1:$\\\rule{0ex}{4.45in}}}
\llap{\parbox[b]{6.25in}{\raggedright \small 4\%~~~5\%~~~6\%~~~7\%~~~8\%~~~9\%~~10\%~11\%~~12\%~~13\%~14\%~~13\%~~12\%~11\%~~10\%~~~9\%~~~8\%~~~7\%~~~6\%~~~5\%\\\rule{0ex}{4.3in}}}
\llap{\parbox[b]{13.2in}{\small (a)\\\vspace{1.9in}(b)\\\vspace{1.95in}(c)\\\rule{0ex}{2.4in}}}
\caption{\label{fig:lams_iz18_nopre} Same as Fig.~\ref{fig:lams_iz13_nopre} except that all quantities are evaluated at 41~km.}
\end{figure*}

After discussing the impact of FFC on the overall evolution of neutrino number densities, we now examine the full time evolution of several key quantities at two representative radii at $r=36$~km (Fig.~\ref{fig:lams_iz13_nopre}) and $41$~km (Fig.~\ref{fig:lams_iz18_nopre}), which are located inside and near the $\nu_e$ sphere at 40~km.   

Fig.~\ref{fig:lams_iz13_nopre}(a) and (b) show the evolution of $I_>$ (green curves), $I_<$ (orange curves), and $\eta_{\rm osc}$ (black curves), which are key quantities characterizing the E-XLN crossings at $r=36$~km. 
Note that $I_<$ only exists when the E-XLN crossing is present. 
The gray vertical lines in all panels separate different stages, whose $b_1$ values are indicated in the middle panel. 
Quantities obtained in fiducial QKE models that adopt different values of $a_1$ are shown by the solid, dashed, and dotted curves for QKE-H, QKE-M, and QKE-L, respectively. 
For the model without including flavor conversions, the values at the end of each stage are shown by the star symbols. 
The evolution of the number density of $\nu_e$, $\bar\nu_e$, $\nu_x$, and $\bar\nu_x$ is shown by the blue, red, cyan, and magenta curves or symbols in panel (c).

The star symbols in panel (c) show again that in the absence of flavor oscillations, $n_{\nu_e}$ decreases and $n_{\bar\nu_e}$ increases when $b_1$ is increased.
At this radius, the E-XLN crossing first appears at $t\gtrsim 0.32$~ms after $b_1$ is switched to 0.06, indicated by the appearance of $I_<$ (orange lines) in panel (b).
Without oscillations, the E-XLN crossing becomes deeper as shown by the orange stars in panel (b).
However, when FFC is turned on, $|I_<|$ is largely suppressed relative to $|I_>|$, since FFC tends to remove the coarse-grained E-XLN crossings such that $I_<$ appears nearly constant in the scale shown in panel (b).
Similarly, $\eta_{\rm osc}$ (black lines) also stays close to zero in panel (b) due to the same reason. 

When zooming in by about a factor of 100, panel (a) reveals that there actually exist three different features during the evolution after the E-XLN crossings appear.
Focusing on the QKE-H model (solid curves), panel (a) shows that while $|I_<|\ll 10^{30}$~cm$^{-3}$ most of the time, reflecting that the system mostly evolves in a near-quasistationary fashion, i.e., the crossings are nearly vanishing, sudden changes of $I_<$ happen around $1.2$~ms and $2.0$~ms, highlighted by the gray bands. 
These sudden changes are related to the transition from $n_{\nu_e}>n_{\bar\nu_e}$ to $n_{\nu_e}<n_{\bar\nu_e}$, which occurs at $t\simeq 1.2$~ms when oscillations are included [see panel (c)], and from $n_{\nu_e}<n_{\bar\nu_e}$ to $n_{\nu_e}>n_{\bar\nu_e}$ at $\simeq 2.0$~ms. 
These ``\emph{swapping}'' episodes correlate with the flavor swap feature discussed above [see Fig.~\ref{fig:rn_overall}(d)], and are related to the overall sign change of E-XLN and $I_>$ [see panel (b)]. 
Accordingly, $I_<$ also changes signs before and after the swapping as clearly shown by panel (a). 

During the times when $|I_<|$ is very small, the evolution of the system can be further separated into two different kinds, depending on the value of $\eta_{\rm osc}$. 
For episodes at $0.8$~ms$\lesssim t\lesssim 1.12$~ms, $1.28$~ms$\lesssim t\lesssim 1.92$~ms, and $2.12$~ms$\lesssim t\lesssim 2.56$~ms, $\eta_{\rm osc}$ is negative with $|\eta_{\rm osc}|$ comparable to $|I_<|$, signifying nearly perfect elimination of the coarse-grained E-XLN crossing. 
Thus, we call these periods ``\emph{near-crossing-elimination}'' episodes.
In contrast, when $0.32$~ms$\lesssim t\lesssim 0.8$~ms and $2.56$~ms$\lesssim t\lesssim 2.72$~ms highlighted by purple bands, $|\eta_{\rm osc}|$ still takes substantially larger values and flips between negative and positive, suggesting that either shallow crossing(s) remain or coarse-grained flavor overconversion occurs.  
For simplicity, we use the term ``\emph{shallow-crossing}'' episodes for these phases.

In Fig.~\ref{fig:lams_iz18_nopre}, we show the evolutions of the same quantities at $r=41$~km. 
The evolution can be similarly separated into shallow-crossing episodes, near-crossing-elimination episodes, and swapping episodes.
The main difference is that FFC starts to operate earlier at $\sim 0.02$~ms, and the flavor swap phenomena also appear at different times. 
Interestingly, the dynamical evolutions during the two swapping episodes are more distinct at this radius than at 36~km, suggesting a substantial break of time-reversal symmetry. 
We will discuss this phenomenon further in Sec.~\ref{subsec:swapping}.
For completeness, we show the evolution of the same quantities at $r=50$~km in Fig.~\ref{fig:lams_iz25_nopre} in Appendix~\ref{sec:A_time_evolution}.

Comparing QKE models that take different attenuation values of $a_1$, panels (b) and (c) of Figs.~\ref{fig:lams_iz13_nopre}, \ref{fig:lams_iz18_nopre}, and \ref{fig:lams_iz25_nopre} show that they give rise to qualitatively similar results, demonstrating that the evolution does not sensitively depend on the choice of the attenuation factor for the forward scattering potential. 
Slightly larger deviations of $I_<$ and $\eta_{\rm osc}$ from zero for models adopting smaller $a_1$ (more attenuated) shown in panel (a) of these figures further indicate that for the unattenuated case, its evolution should be closer to near-quasistationary, confirming the finding of Ref.~\cite{xiong2025robust}. 
Detailed discussions on the impact of the artificial attenuation will be given in Sec.~\ref{sec:attenuation}.

Note that during the shallow-crossing and near-crossing-elimination episodes, the evolution of quantities such as $I_>$ and neutrino number densities of different species tends to be rather smooth and monotonic from one stage to the next. 
However, during certain stages, e.g., between $1.6$~ms and $1.92$~ms, the flavor evolution behaves differently and shows distinct bumps in $I_>$.
We find that this phenomenon is related to the onset of collisional flavor instability~\cite{johns2023collisional,padilla2022neutrino1,johns2022collisional,xiong2023evolution,lin2023collisioninduced,xiong2023collisional,kato2023flavor,liu2023systematic,akaho2024collisional,liu2023universality,kato2024collisional,fiorillo2024collisions,shalgar2024neutrinos}. 
These features suggest that collisional effects can, under certain conditions, contribute non-negligibly to the flavor evolution.
However, as the primary focus of this work is on FFCs, a systematic investigation of collisional flavor instability and its interplay with FFC will be deferred to future work.

\section{Dynamics of the three evolution episodes}
\label{sec:episodes}
In this section, we analyze in detail the dynamics of the three evolution episodes identified above.
Sec.~\ref{subsec:shallow} examines the onset of FFCs from the emergence of shallow E-XLN crossings, Sec.~\ref{subsec:elimination} the near-crossing-elimination episode together with the role of collisional decoherence in sustaining it, and Sec.~\ref{subsec:swapping} the swapping episode characterized by a dynamically propagating E-XLN zero surface.

\subsection{Dynamics of FFC onset and shallow-crossing episode}
\label{subsec:shallow}

\begin{figure*}
    \centering
    \includegraphics[width=0.95\linewidth]{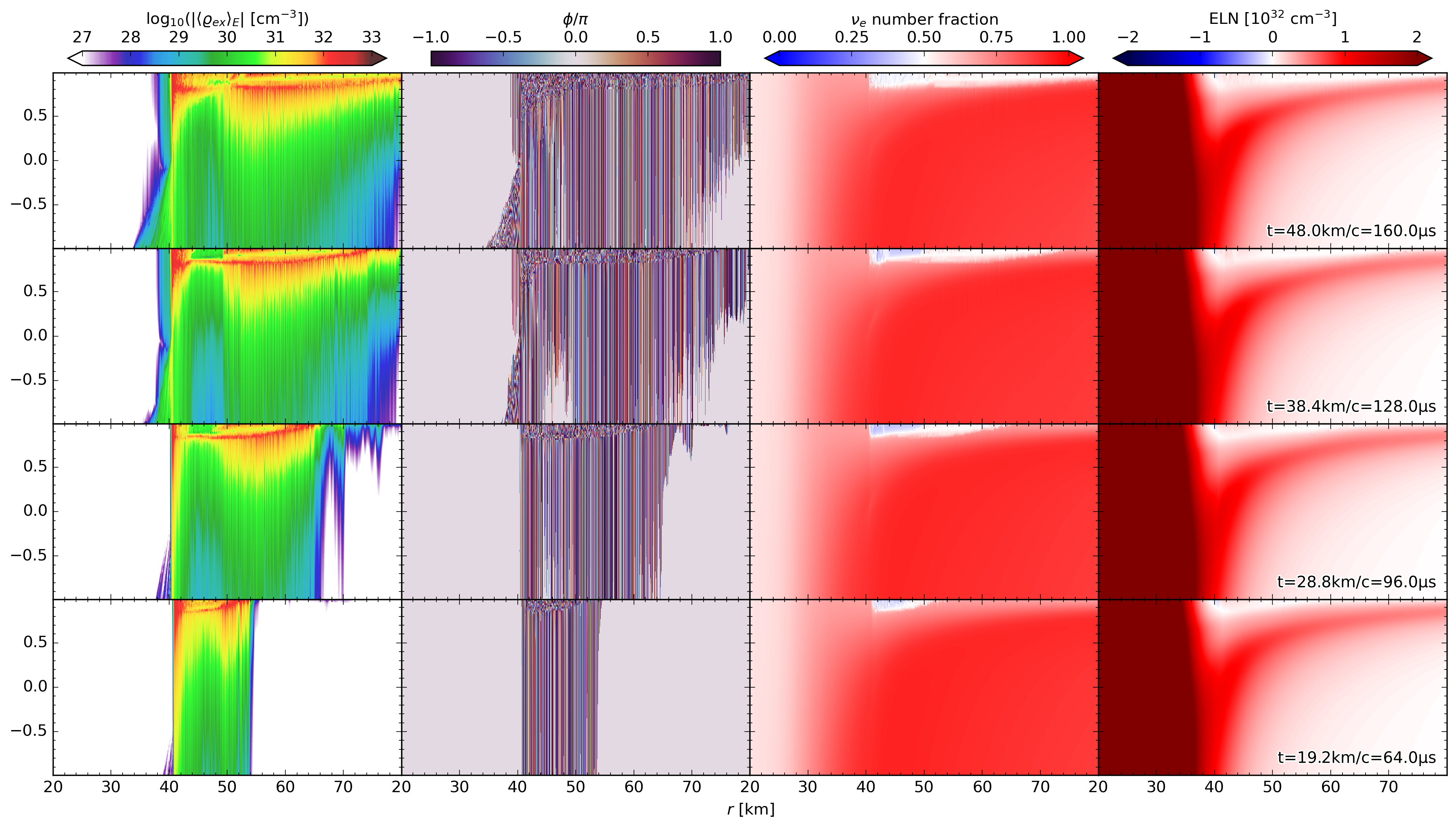}
    \caption{Evolution of the off-diagonal flavor coherence $\langle\varrho_{ex}\rangle_E$ (first column), the associated phase angle $\phi_{ex}$ (second column), the $\nu_e$ number fraction (third column), and E-XLN angular distribution (fourth column) in $20$~km~$<r<80$~km for Model QKE-H. Y-axis is for $v_r$.}
    \label{fig:details_stage_1}
\end{figure*}

Fig.~\ref{fig:details_stage_1} shows the radial and velocity profiles of the off-diagonal density matrix element $\langle\varrho_{ex}\rangle_E$ (Eq.~\eqref{eq:avgrho}), 
the phase angle of the off-diagonal component $\phi_{ex}$ (Eq.~\eqref{eq:phi_emu}), the $\nu_e$ number fraction (Eq.~\eqref{eq:chi}), and the (non-coarse-grained) E-XLN angular distribution $G$ for five snapshots
during the first 0.16~ms in the QKE-H model.
These panels show that a very shallow E-XLN crossing first appears at $\lesssim 40$~km in the forward-propagating direction $v_r\gtrsim 0.8$; see the bottom-rightmost panel.
Note that substantial FFCs develop at $r\gtrsim 40$~km only when the crossing becomes slightly deeper such that the FFI growth rate is large enough.

The FFCs progressively influence the outer region as flavor waves propagate and lead first to slight flavor overconversions in the forward direction (blue patches in $\nu_e$ number fraction panels), which later settle into near-flavor-equipartition (white patches in $\nu_e$ number fraction panels). 
Similar to results obtained in two-step models in Ref.~\cite{xiong2024fast}, the phase angle $\phi_{ex}$ shows small-scale and incoherent (coherent) features in the $v_r$ domain (un)affected by FFCs.
These structures clearly indicate that FFCs associated with shallow E-XLN crossings carry characteristic length scales, similar to those generated by deep crossings in two-step models~\cite{xiong2024fast}.
The spectrogram analysis for $\varrho_{ex}$ as done in Ref.~\cite{xiong2024fast}, which will be discussed in Sec.~\ref{sec:spectrogram} in detail, shows that the dominant wave number ranges from $\simeq -300$~km$^{-1}$ at $r\simeq 45$~km to $\simeq -50$~km$^{-1}$ at $r\simeq 80$~km during this episode in the QKE-H model (see the bottom right panel of Fig.~\ref{fig:FFT_LSA}). 
This implies that it is important to have enough spatial resolution to capture the dynamics of FFCs in global simulations where E-XLN crossings gradually appear, consistent with the suggestions in Ref.~\cite{xiong2024fast}.

\subsection{Near-crossing-elimination episode}
\label{subsec:elimination}

\begin{figure}
    \centering
    \includegraphics[width=\columnwidth]{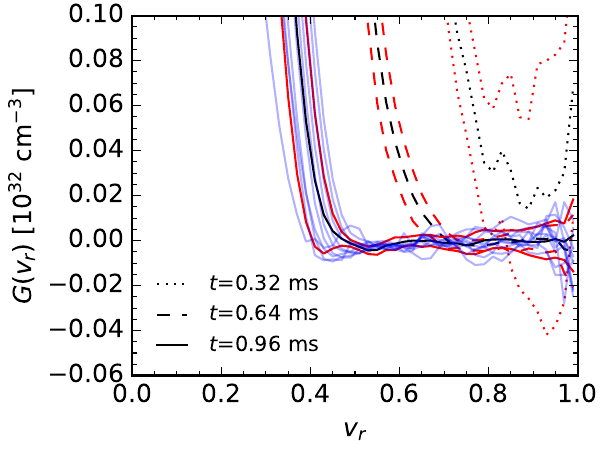}
    \caption{\label{fig:G_time_dependent}
    E-XLN angular distributions at $r=41$~km in Model QKE-H at the end of three stages, $t=0.32$, $0.64$, and $0.96$~ms (dotted, dashed, and solid curves, respectively).
    Black curves show the coarse-grained distributions $\tilde G(v_r)$, and red curves the $1\sigma$ range of all $G(v_r)$ within the subgrid of width $\Delta r=1.2$~km, while the thin blue curves show a few selected $G(v_r)$ at $t=0.96$~ms.
    As the system evolves through successive stages, the E-XLN crossing progressively moves toward lower $v_r$, and an increasingly wider velocity range on the smaller side becomes very flat.}
\end{figure}

As the system evolves beyond the shallow-crossing episode, FFCs begin to have a more pronounced impact on the E-XLN angular distribution.
The continued action of FFCs drives the system toward a state where the crossing becomes significantly weakened, marking the transition into the near-crossing-elimination episode.

In this episode, the smaller side of the E-XLN becomes flatter and shallower.
This behavior is illustrated in Fig.~\ref{fig:G_time_dependent}, which shows the E-XLN angular distributions at $r=41$~km, near the $\nu_e$ sphere, at the end of three different stages at $t=0.32$, 0.64, and 0.96~ms.
The black curves show the coarse-grained distributions $\tilde G$, while the red curves indicate the corresponding $1\sigma$ range of all $G$ within the subgrid of width $\Delta r=1.2$~km, quantifying the small-scale variations.
At $t=0.32$~ms, the system is still in the shallow-crossing episode, and the distribution on the smaller side exhibits large variations around $\tilde G$.
At the two later times, the system has entered the near-crossing-elimination episode, and the variations become much smaller.
Meanwhile, as the local conditions evolve, the crossing of $\tilde G$ progressively moves toward lower $v_r$, while the distribution over an increasingly wider velocity range on the smaller side is kept very flat and close to zero by the continued action of FFCs, in contrast to the deepening crossings that would develop in the absence of oscillations (see the star symbols in Fig.~\ref{fig:lams_iz18_nopre}).

In this episode, the flavor conversion reflects a dynamically regulated balance between oscillations and collisional effects, with persisting marginal instabilities (see Sec.~\ref{sec:spectrogram}).
The balance is regulated by adjusting $|I_<|$ within a range that prevents it from becoming too large, which would otherwise trigger more violent flavor conversion.
If $|I_<|$ were significantly larger, the growth rate of FFCs would increase sharply, leading to rapid depletion of the E-XLN and potentially overshooting a stable configuration.
At the same time, $|I_<|$ cannot be too small, because then the flavor conversion is delayed and the E-XLN angular distribution evolves mainly due to collisions rather than coherent oscillations.
This interplay continues until the growth rate of the instability roughly balances the collisional rate, such that $|I_<|$ self-regulates at a moderate level.

\subsubsection{Role of collisions in decoherence}
\label{sec:decoherence}
A more fundamental ingredient underlying this balance, which is also related to the success of ECT models that will be discussed in Sec.~\ref{sec:comparison}, is collisional decoherence.
In what follows, we discuss how collisional decoherence facilitates this balance by effectively regulating the off-diagonal elements.

\begin{figure*}
    \centering
    \includegraphics[width=0.95\columnwidth]{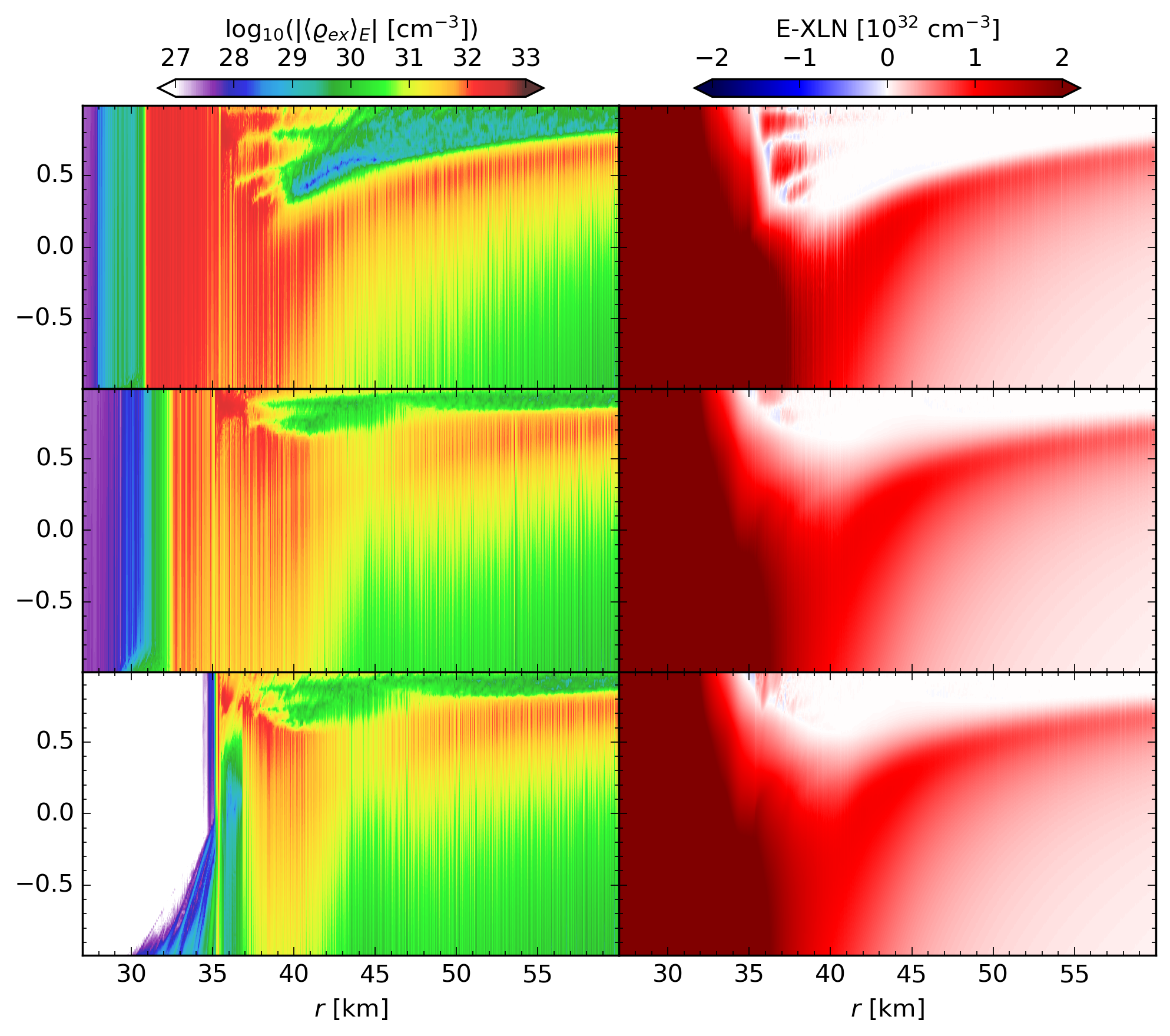}
		\llap{\parbox[b]{1.in}{\small no collisional decoherence,\\+ no NNS\\\vspace{.49in}no collisional decoherence\\\vspace{.49in}QKE-H\\$t=0.64$~ms\\\rule{0ex}{0.3in}}}
    \includegraphics[width=0.95\columnwidth]{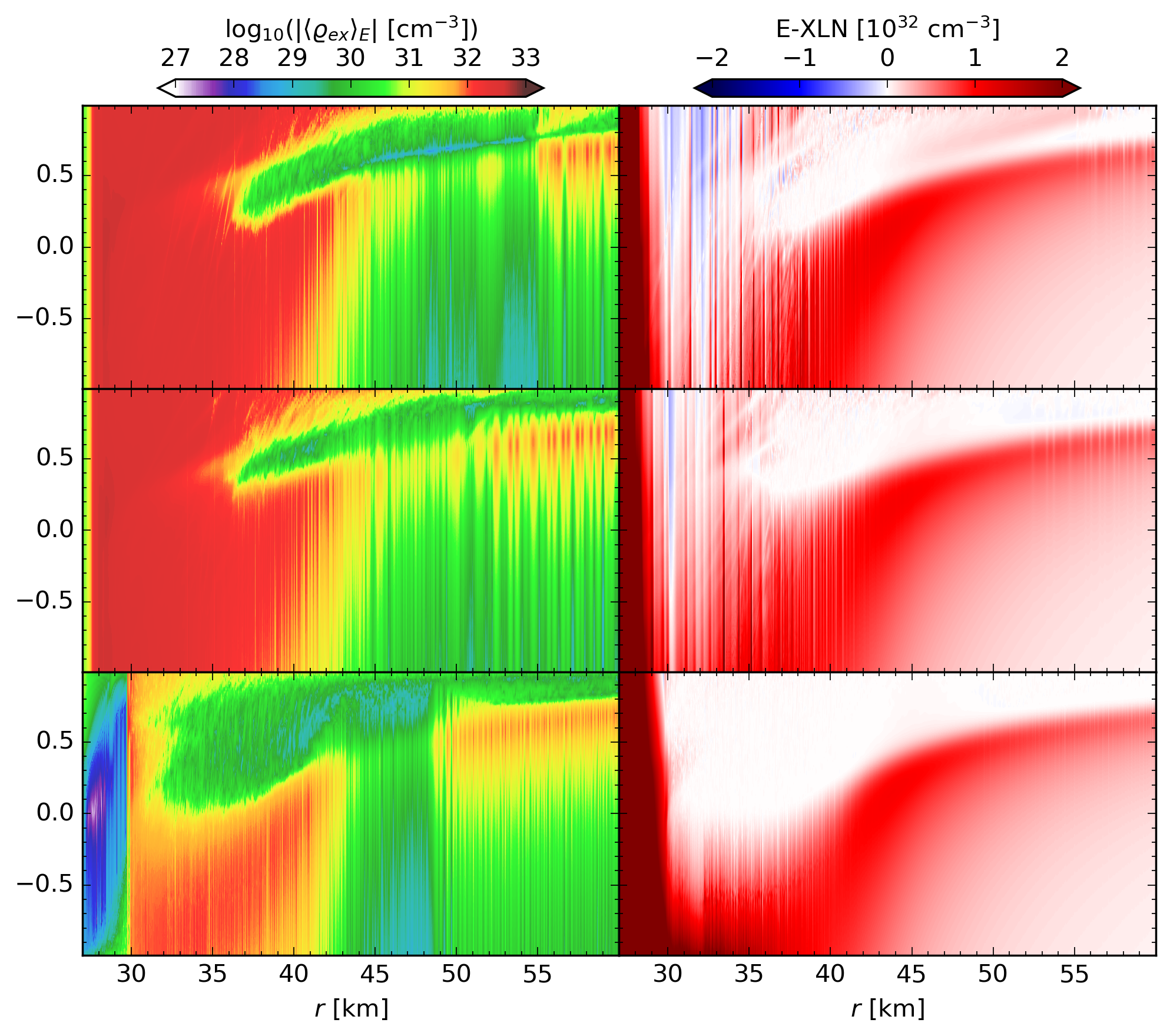}
		\llap{\parbox[b]{1.in}{\small no collisional decoherence,\\+ no NNS\\\vspace{.49in}no collisional decoherence\\\vspace{.49in}QKE-L\\$t=1.12$~ms\\\rule{0ex}{0.3in}}}
    \caption{Snapshots of the off-diagonal flavor coherence $\langle\varrho_{ex}\rangle_E$ (first column) and E-XLN angular distribution (second column) in $30$~km~$<r<60$~km for two additionally evolved scenarios (top and middle rows) based on the snapshots of Model QKE-H at $t=0.64$~ms (third row, left panels) and Model QKE-L at $t=1.12$~ms (bottom row, right panels). 
    Y-axis is for $v_r$.
    For both scenarios, collisional decoherence is excluded for the whole radial domain.  
    In the first scenario (top row), the NNS process is additionally excluded for $r>35$~km.
    }
    \label{fig:decoherence}
\end{figure*}

In addition to the long-term time-dependent models listed in Table~\ref{tab:parameters}, we consider two more contrasting scenarios to isolate and examine the effects of collisional decoherence.
We select the time $t = 0.64$~ms for Model QKE-H and $t = 1.12$~ms for Model QKE-L as the starting points for the new simulations.
The corresponding off-diagonal flavor coherence and E-XLN distribution are shown in the bottom panels of Fig.~\ref{fig:decoherence}.
The background matter profiles, together with the derived collisional rates, are then fixed and held stationary at their values corresponding to this time.
We then resume the evolution for an additional duration of 0.16~ms in two scenarios: one excludes the off-diagonal collisional components over the whole radial range, while the other additionally excludes the NNS process for $r>35$~km.

The first two rows of Fig.~\ref{fig:decoherence} show the results obtained at the end of these additional simulations.
For the models derived from Model QKE-H, in the scenario without collisional decoherence (middle panel), the E-XLN distributions remain essentially unchanged.
However, the magnitude of the off-diagonal element $\langle \varrho_{ex} \rangle_E$ inside the neutrinosphere is significantly larger than in the case where damping is included.
This suggests that collisional effects play a critical role in suppressing coherent flavor mixing and limiting the growth of off-diagonal correlations.

For the scenario without collisional decoherence and without NNS above $35$~km,  
the E-XLN distributions at $\sim 35$--$40$~km are substantially affected, showing a larger amount of positive E-XLN in forward-propagating directions ($v_r>0$), accompanied by slightly larger values of $\langle \varrho_{ex} \rangle_E$ and small-scale fluctuations.
This behavior arises because scattering between neutrinos and nucleons tends to smooth the angular distributions.
A neutrino propagating along a given direction becomes mixed with neutrinos arriving from other directions, so that the variations of E-XLN among different angles are gradually reduced. 
For $\langle \varrho_{ex} \rangle_E$, including NNS effectively introduces additional kinematic decoherence arising from the dephasing among neutrinos propagating along different trajectories \cite{raffelt2007selfinduced,johns2020fast,bhattacharyya2021fast,xiong2023symmetry,liu2025dynamical}. 
Thus, $\langle \varrho_{ex} \rangle_E$ becomes larger when NNS is excluded. 

For the models derived from Model QKE-L (right two columns in Fig.~\ref{fig:decoherence}), removing collisional decoherence, either alone or together with NNS, has an even larger impact.
The off-diagonal element $\langle \varrho_{ex} \rangle_E$ grows to even larger values and propagates outward, indicating that the system no longer remains close to the marginal configuration maintained by collisional decoherence. 
Consequently, the angular distributions on the small side are no longer flattened and smoothed effectively.
The E-XLN angular crossing therefore can persist rather than being eliminated.

\subsection{Swapping episode}
\label{subsec:swapping}

\begin{figure}
    \centering
    \includegraphics[width=0.98\columnwidth]{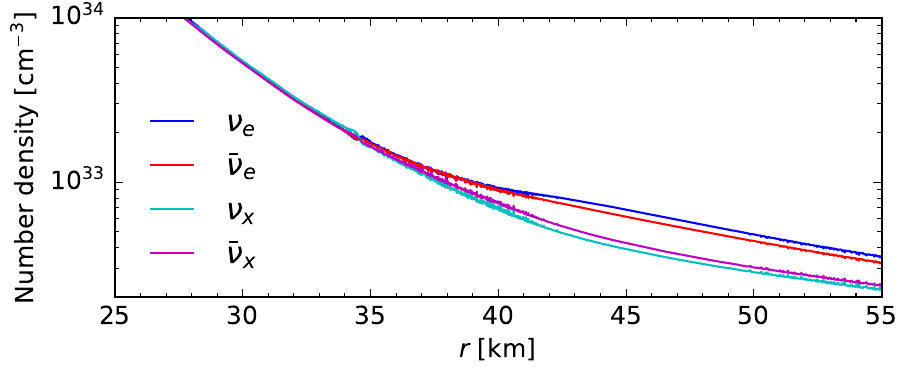}
\llap{\parbox[b]{1.2in}{\small $t=1.152$~ms\\\rule{0ex}{1.1in}}} \includegraphics[width=0.98\columnwidth]{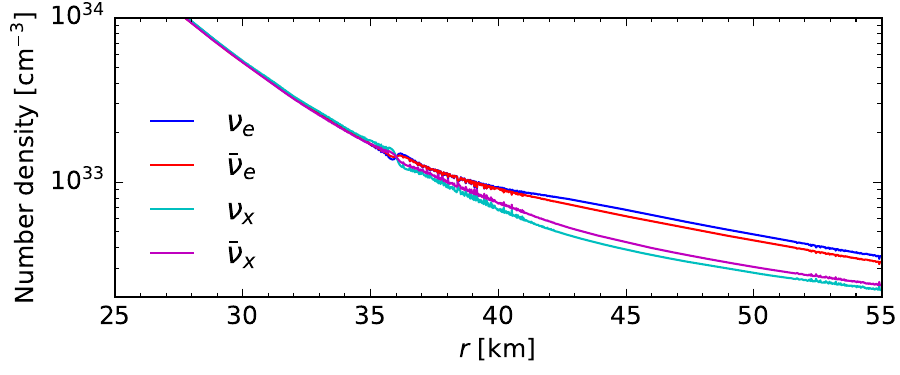}
\llap{\parbox[b]{1.2in}{\small $t=1.184$~ms\\\rule{0ex}{1.1in}}}  \includegraphics[width=0.98\columnwidth]{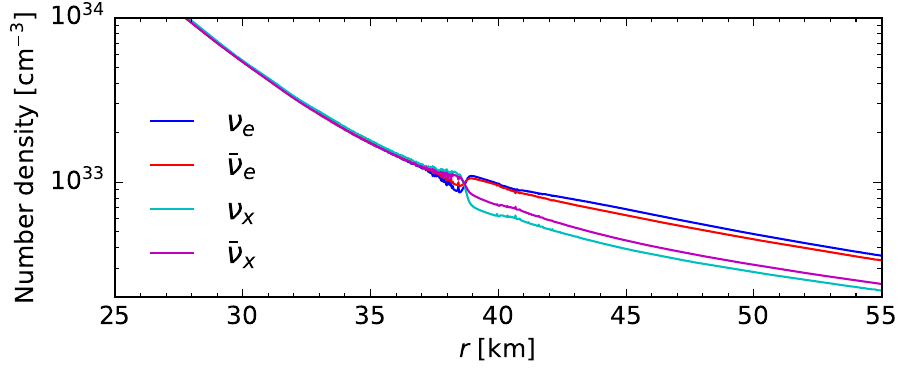}
\llap{\parbox[b]{1.2in}{\small $t=1.28$~ms\\\rule{0ex}{1.1in}}}	 \includegraphics[width=0.98\columnwidth]{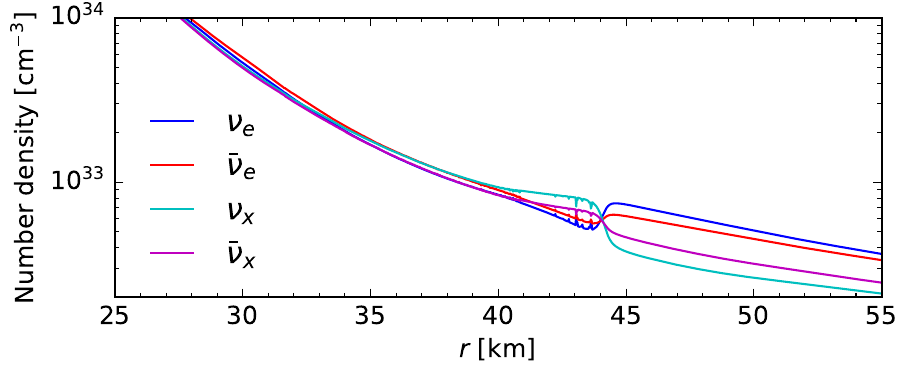}
\llap{\parbox[b]{1.2in}{\small $t=1.36$~ms\\\rule{0ex}{1.1in}}}	  \includegraphics[width=0.98\columnwidth]{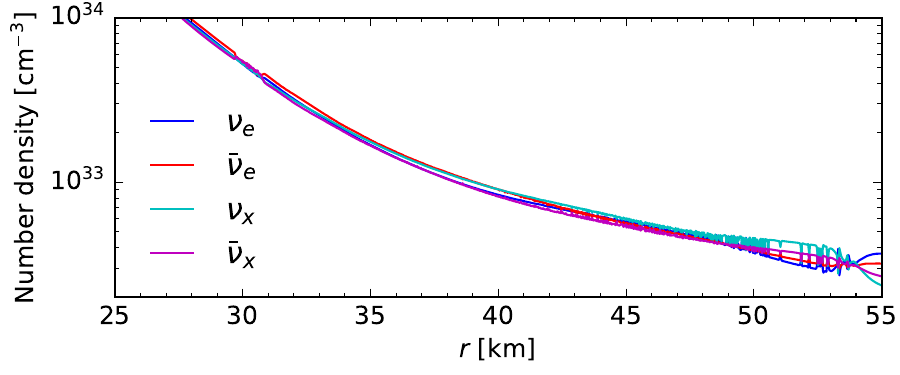}
\llap{\parbox[b]{1.2in}{\small $t=1.44$~ms\\\rule{0ex}{1.1in}}}
    \caption{Radial profiles of neutrino number densities between $r=25$ and 55 km at different times during the swapping episode in model QKE-H.}
    \label{fig:swapping}
\end{figure}

\begin{figure}
    \centering
    \includegraphics[width=0.98\columnwidth]{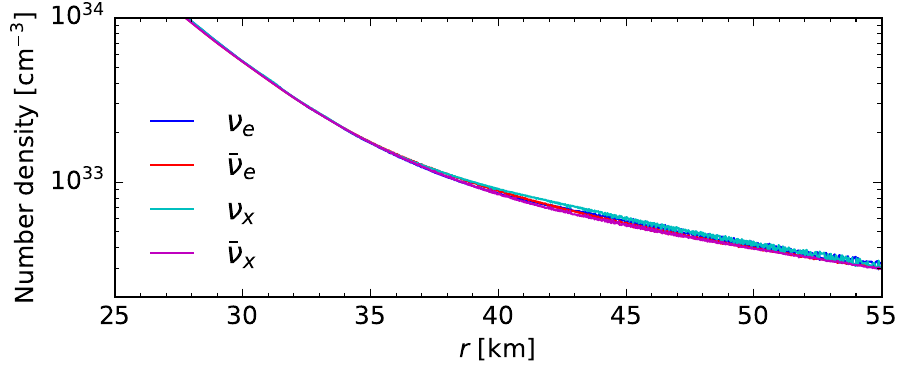}
\llap{\parbox[b]{1.2in}{\small $t=1.792$~ms\\\rule{0ex}{1.1in}}} \includegraphics[width=0.98\columnwidth]{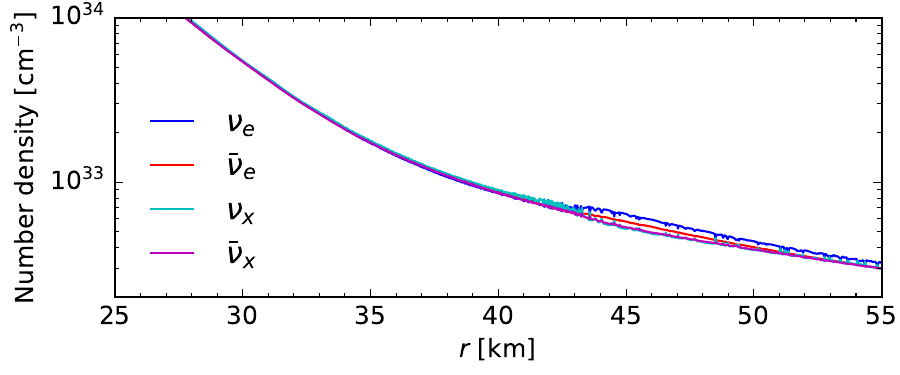}
\llap{\parbox[b]{1.2in}{\small $t=1.856$~ms\\\rule{0ex}{1.1in}}} \includegraphics[width=0.98\columnwidth]{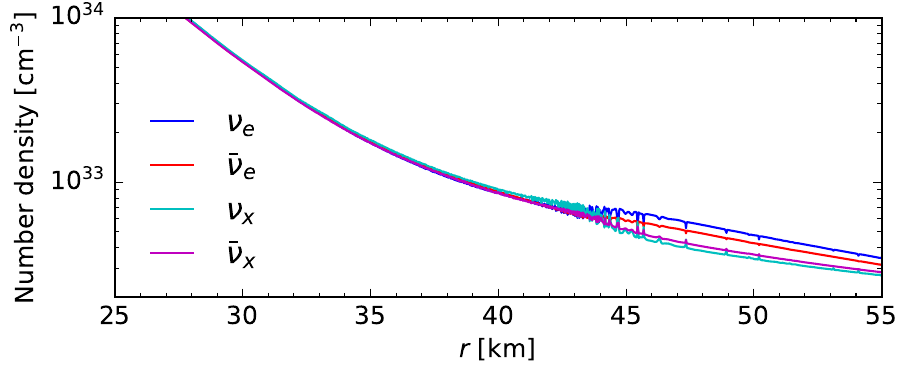}
\llap{\parbox[b]{1.2in}{\small $t=1.92$~ms\\\rule{0ex}{1.1in}}} \includegraphics[width=0.98\columnwidth]{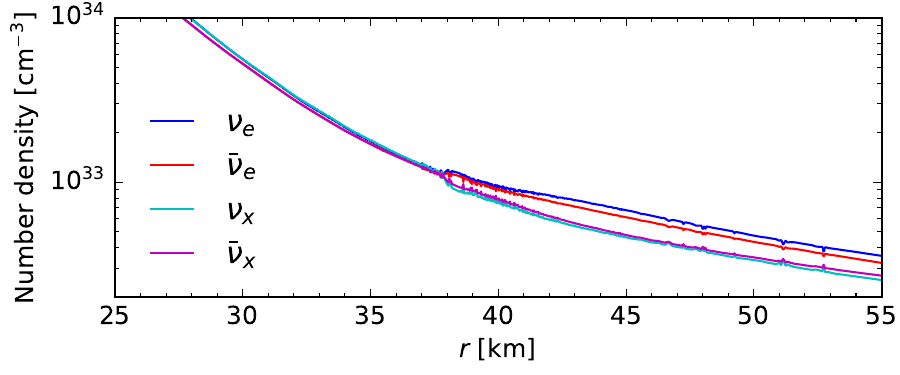}
\llap{\parbox[b]{1.2in}{\small $t=2.0$~ms\\\rule{0ex}{1.1in}}}	 \includegraphics[width=0.98\columnwidth]{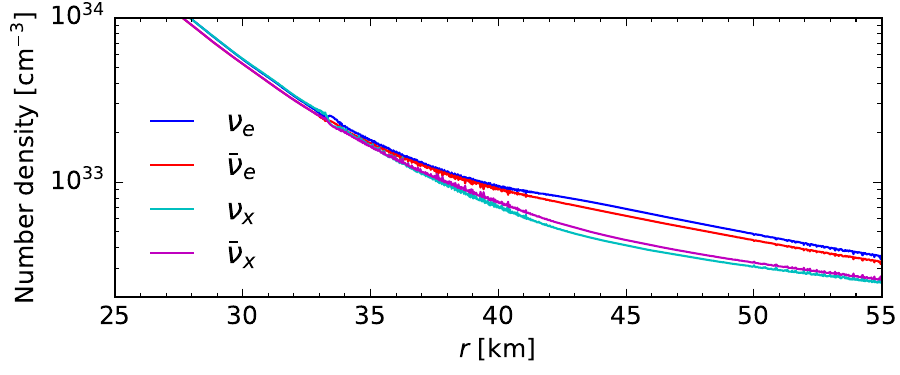}
\llap{\parbox[b]{1.2in}{\small $t=2.08$~ms\\\rule{0ex}{1.1in}}}
    \caption{Similar to Fig.~\ref{fig:swapping} but during the reversal phase of the swapping episode.}
    \label{fig:time_reversal}
\end{figure}

\begin{figure}
    \centering
    \includegraphics[width=0.95\columnwidth]{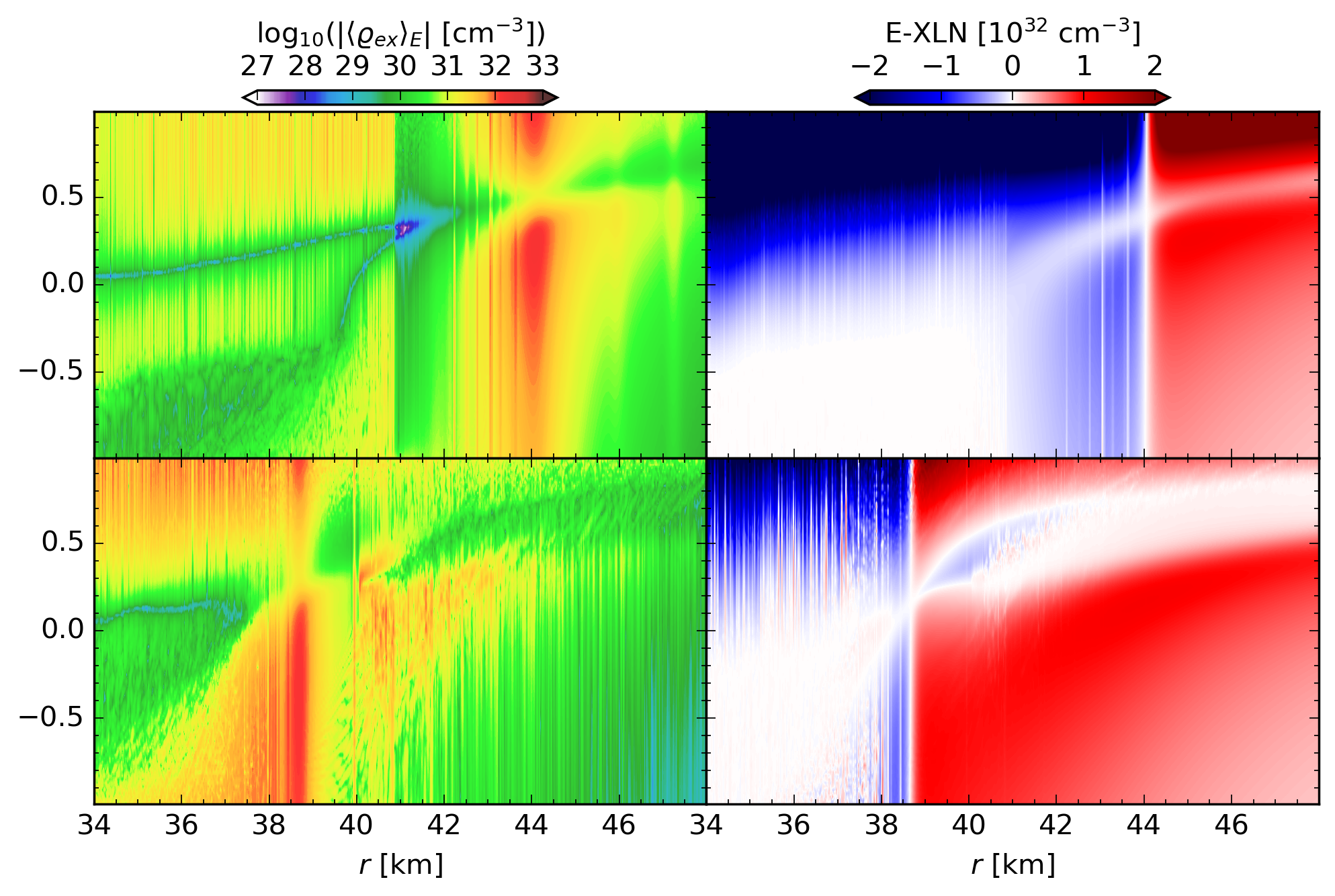}
		\llap{\parbox[b]{1.in}{\small $t=1.36$~ms\\\vspace{.7in}$t=1.28$~ms\\\rule{0ex}{0.3in}}}
    \caption{Snapshots of the off-diagonal flavor coherence $\langle\varrho_{ex}\rangle_E$ (first column) and E-XLN angular distribution (second column) in Model QKE-H at $t=1.28$~ms (lower panel) and at $t=1.36$~ms (upper panel). Y-axis is for $v_r$.
    }
    \label{fig:swapping_2d}
\end{figure}

As already discussed in Sec.~\ref{sec:global}, the increasingly neutron-rich matter background leads to a reduced ratio of $n_{\nu_e}/n_{\bar\nu_e}$ due to the change of weak-interaction balance condition, driving the E-XLN toward zero.
When oscillations are included, the E-XLN crosses zero first around $t\sim 1.1$~ms at $r\sim 35$~km, shown by the top panel of Fig.~\ref{fig:swapping}, which displays the neutrino number density of different species between $t=1.152$~ms and 1.44~ms.
These panels show that the transition from positive to negative E-XLN at inner radii results in the launch of an E-XLN zero surface that moves outward over time.

The presence of the zero surface drastically changes the behavior of flavor conversions.
Fig.~\ref{fig:swapping} shows that outside the zero surface, $n_{\nu_e}>n_{\nu_x}$ and $n_{\bar\nu_e}>n_{\bar\nu_x}$.
However, these orderings are flipped inside the surface where $n_{\nu_e}<n_{\nu_x}$ and $n_{\bar\nu_e}<n_{\bar\nu_x}$. 
These features suggest that nearly complete flavor swap happens across the zero surface, similar to the phenomenon discussed in \cite{nagakura2023global,zaizen2024fast}.
In fact, $|\langle \varrho_{ex}\rangle_E|$ and the E-XLN distribution as functions of $v_r$ and $r$ shown in
Fig.~\ref{fig:swapping_2d} at $t=1.28$~ms and 1.36~ms clearly demonstrate a $v_r$-independent surface over which large $\varrho_{ex}$ exists, which facilitates nearly complete flavor conversion for neutrinos crossing the surface~\cite{zaizen2024fast}.

One interesting and distinctive aspect of the swapping behavior in our time-dependent simulation is that the E-XLN zero crossing dynamically propagates rather than being fixed in space.
Such mobility of the zero surface naturally arises in our setup because neutrinos are produced self-consistently through collisions instead of being injected from fixed boundaries, unlike the setup in \cite{nagakura2023global,zaizen2024fast}.
As a result, the zero surface in our simulation is not anchored at a single radius but propagates as the local balance between $\nu_e$ and $\bar{\nu}_e$ changes.

The dynamical evolution of the reverse swapping episode that occurs during the second half of the evolution, when $b_1$ is decreased back and the system returns to the $\nu_e$-dominant conditions, is shown in Fig.~\ref{fig:time_reversal}.
In this case, the swapping surface forms at approximately $43~\text{km}$ around $t=1.856$~ms, outside which more $\nu_e$'s are first produced, followed by also enhanced $\bar\nu_e$ at slightly later snapshots.
Differently from the first phase, the zero surface is less pronounced and moves inward, reaching $\sim 38$~km at $t=2.0$~ms before it disappears inside the neutrinosphere. 

The different behavior in the two swapping episodes thus reveals a clear breaking of the time-reversal symmetry (also shown in Figs.~\ref{fig:lams_iz18_nopre} and \ref{fig:lams_iz25_nopre}). 
If the evolution of the system can always be characterized by near-quasistationary states that change adiabatically in time, one would naively expect nearly identical evolution for the stages with the same background profiles (same $b_1$ values), independently of whether $b_1$ increases or decreases.
The occurrence of swaps here highlights the dynamical and out-of-equilibrium nature of the problem for a system that evolves in time. 

\section{Spectrogram and linear stability analysis}
\label{sec:spectrogram}
In this section, we quantify the characteristic length scales of the flavor waves developed during the different episodes using the spectrogram analysis, and interpret them with the linear stability analysis.

Following our previous work \cite{xiong2024fast}, we perform the spectrogram analysis to quantify the dominant length scale associated with flavor waves obtained in numerical simulations.
For a given time snapshot, we partition the whole simulation radial domain into $N_r^{\rm sp}$ blocks and perform discretized Fourier transform for the neutrino off-diagonal mixing within each block centered at a radius $r$ by 
\begin{align}
    & \mathcal F_{r,K_r} = \frac{1}{\Delta r^\mathrm{sp}} \int_{\rm block} dr'\, e^{-i K_r r'} W(r'-r) \times \nonumber\\
    & \int dv_r\, \langle \varrho_{ex} \rangle_E(v_r,r') \mathrm{sgn}[\langle \varrho_{ee} \rangle_E(v_r,r')-\langle \varrho_{xx} \rangle_E(v_r,r')],
\end{align}
where $\Delta r^\mathrm{sp}$ is the width of each block, $K_r$ are discretized values that are the multiples of $\pi/\Delta r^\mathrm{sp}$, and $W(r)$ is the Tukey window function with a cosine lobe.
We take $N_r^{\rm sp}=50$ and $\Delta r^\mathrm{sp}=1.2$~km to ensure sufficient resolution in $K_r$ space.

In addition to spectrogram analysis, since the system stays mostly in the marginally unstable regime, we also perform two different LSA methods based on the coarse-grained properties of the system.
The first one takes the discretized energy and angular grids consistent with the numerical simulations and performs the diagonalization to identify the unstable normal modes.
For details of this procedure, see, e.g., Refs.~\cite{xiong2023evolution,xiong2024fast}.

For the second LSA method, we take the dispersion relation approach adopted in, e.g., \cite{yi2019dispersion,capozzi2019fast}.
Taking the collisionless and massless (i.e. without vacuum term) limit without the angular advection, and assuming the off-diagonal elements are small, Eqs.~\ref{eq:eom_nu} and \ref{eq:eom_nubar} can be linearized as
\begin{align}\label{eq:LEQ}
& \left( \tilde\Omega - \tilde K_r v_r \right)  Q(v_r) \nonumber\\
= & -\sqrt{2}G_F \int dv_r'\,(1- v_r v_r') \tilde G(v_r') Q(v_r'),
\end{align}
where the energy-integrated collective mode $\langle \varrho_{ex} \rangle_E - \langle \bar\varrho_{ex} \rangle_E = \tilde G(v_r) Q(v_r) e^{-i[\Omega t-K_r(r'-r)]}$ (defined over the coarse-grained angular distributions) is characterized by physical $\Omega$ and $K_r$, which are different from the shifted $\tilde \Omega = \Omega-\sqrt{2}G_F I_0$ and $\tilde K_r = K_r-\sqrt{2}G_F I_1$, and $r'$ denotes the spatial dependence locally around $r$ assuming local homogeneity.
Since the R.H.S. is a linear component in the form of $a+b v_r$, the eigenfunction can be rewritten as
\begin{equation}
    Q(v_r) = \frac{a+bv_r}{\tilde v_{\rm ph} - v_r},
\end{equation}
where $\tilde v_{\rm ph}$ is the phase velocity for the shifted dispersion relation.
Plugging it back, the response function for the dispersion relation is obtained as
\begin{equation}
    \begin{pmatrix}
        \tilde K_r & \\ & -\tilde K_r
    \end{pmatrix} \begin{pmatrix} a \\ b \end{pmatrix}
    = - \sqrt{2}G_F \begin{pmatrix}
        P_0(\tilde v_{\rm ph}) & P_1(\tilde v_{\rm ph}) \\ P_1(\tilde v_{\rm ph}) & P_2(\tilde v_{\rm ph})
    \end{pmatrix}\begin{pmatrix} a \\ b \end{pmatrix} ,
\end{equation}
and the determinant of the matrix has to be zero
\begin{equation}
    \left\lVert \begin{pmatrix}
        \tilde K_r & \\ & -\tilde K_r
    \end{pmatrix} + \sqrt{2}G_F \begin{pmatrix}
        P_0(\tilde v_{\rm ph}) & P_1(\tilde v_{\rm ph}) \\ P_1(\tilde v_{\rm ph}) & P_2(\tilde v_{\rm ph})
    \end{pmatrix} \right\rVert = 0,
\end{equation}
where $P_n$ is the principal value (P.~V.) integral with a pole at real $\omega$:
\begin{equation}
    P_n(\omega) = {\rm P.V.}\int_{-1}^1 dv_r\, \frac{v_r^n \tilde G}{\omega-v_r}.
\end{equation}
Given a value of $\tilde v_{\rm ph}$, one can obtain two branches of $\tilde K$ as
\begin{align}\label{eq:K_boundary}
    \tilde K_r(\tilde v_{\rm ph}) = & \frac{\sqrt{2}G_F}{2} \bigg[ P_2(\tilde v_{\rm ph})-P_0(\tilde v_{\rm ph}) \pm  \nonumber\\
    & \sqrt{[P_0(\tilde v_{\rm ph})+P_2(\tilde v_{\rm ph})]^2-4 P_1^2(\tilde v_{\rm ph})} \bigg].
\end{align}
For an angular distribution with only one angular crossing $v_c$, the unstable branch stops at $v_c$ \cite{yi2019dispersion}.
When the angular crossing is shallow, the onset of the unstable branch turns near-luminal and closes at the shallow side with $|v_r|=1$ \cite{fiorillo2024theory1}.
Inserting $\tilde v_{\rm ph}=v_c$ and $\tilde v_{\rm ph}=1$ into Eq.~\eqref{eq:K_boundary} allows to find the physical $K_r$ boundaries of the two unstable branches.
When the E-XLN crossing does not exist or when the small side is not connected to $v_r=1$, we take $v_c=1$ so the corresponding boundary merges with the one associated with $\tilde v_{\rm ph}=1$.
Since we use the collisionless and massless limit, the attenuation can be considered easily by rescaling $K_r$ and $\Omega$ accordingly for both spectrogram and LSA.
An additional comparison study on the difference between these two methods for the shallow-crossing and near-crossing-elimination episodes will be pursued in the future.

\begin{figure*}[!hbt]
\centering
\includegraphics[width=.24\textwidth]{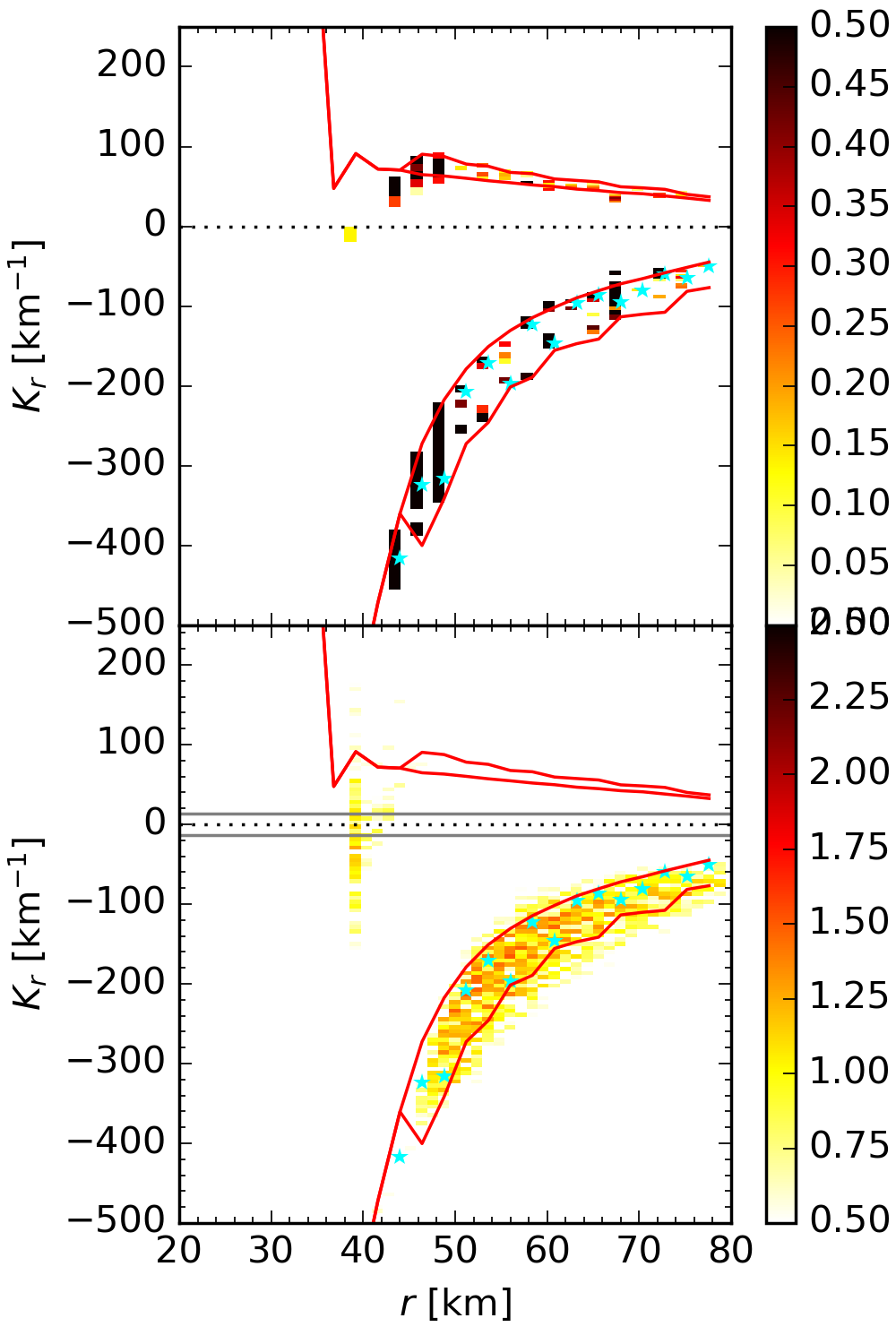}
\includegraphics[width=.24\textwidth]{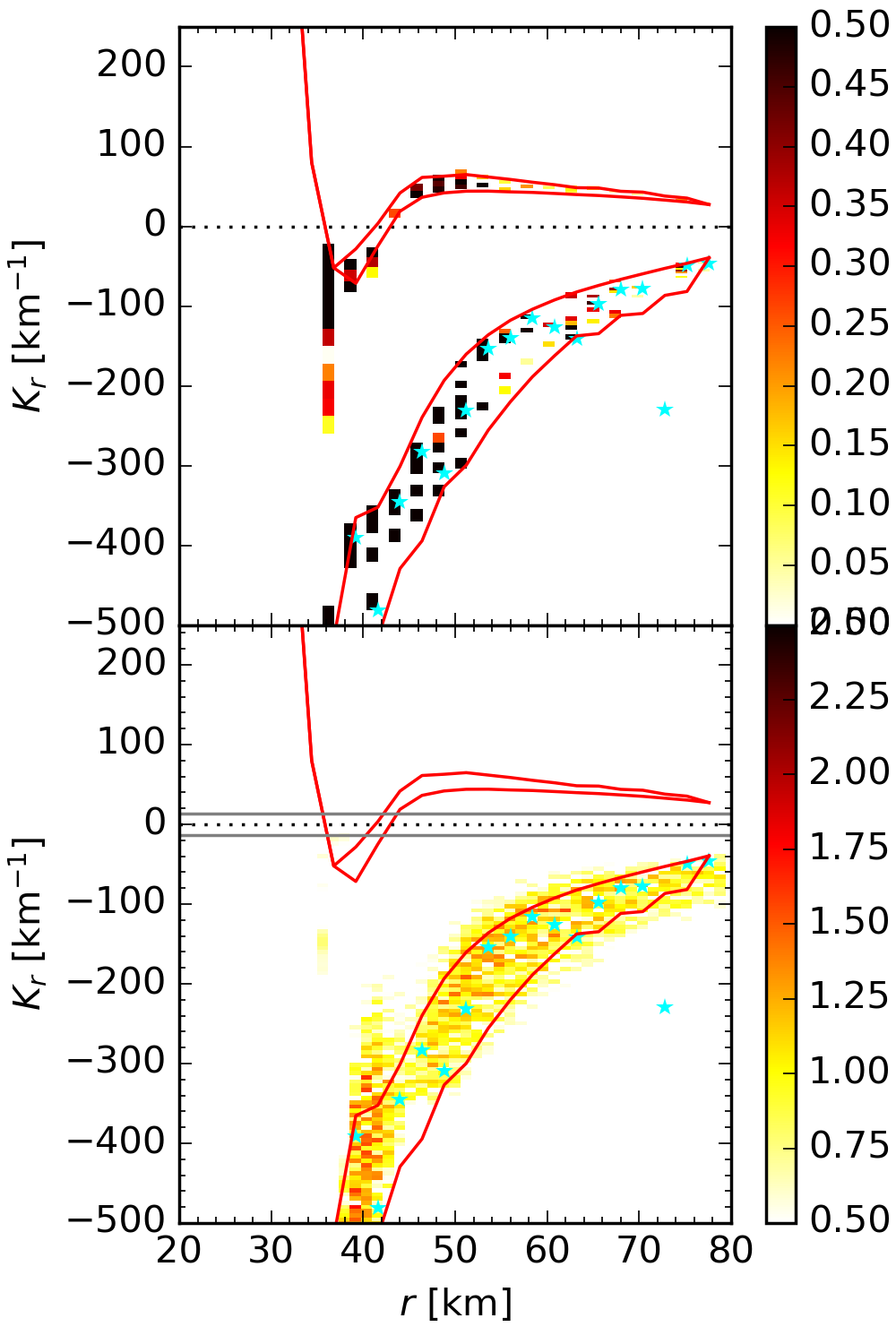}
\includegraphics[width=.24\textwidth]{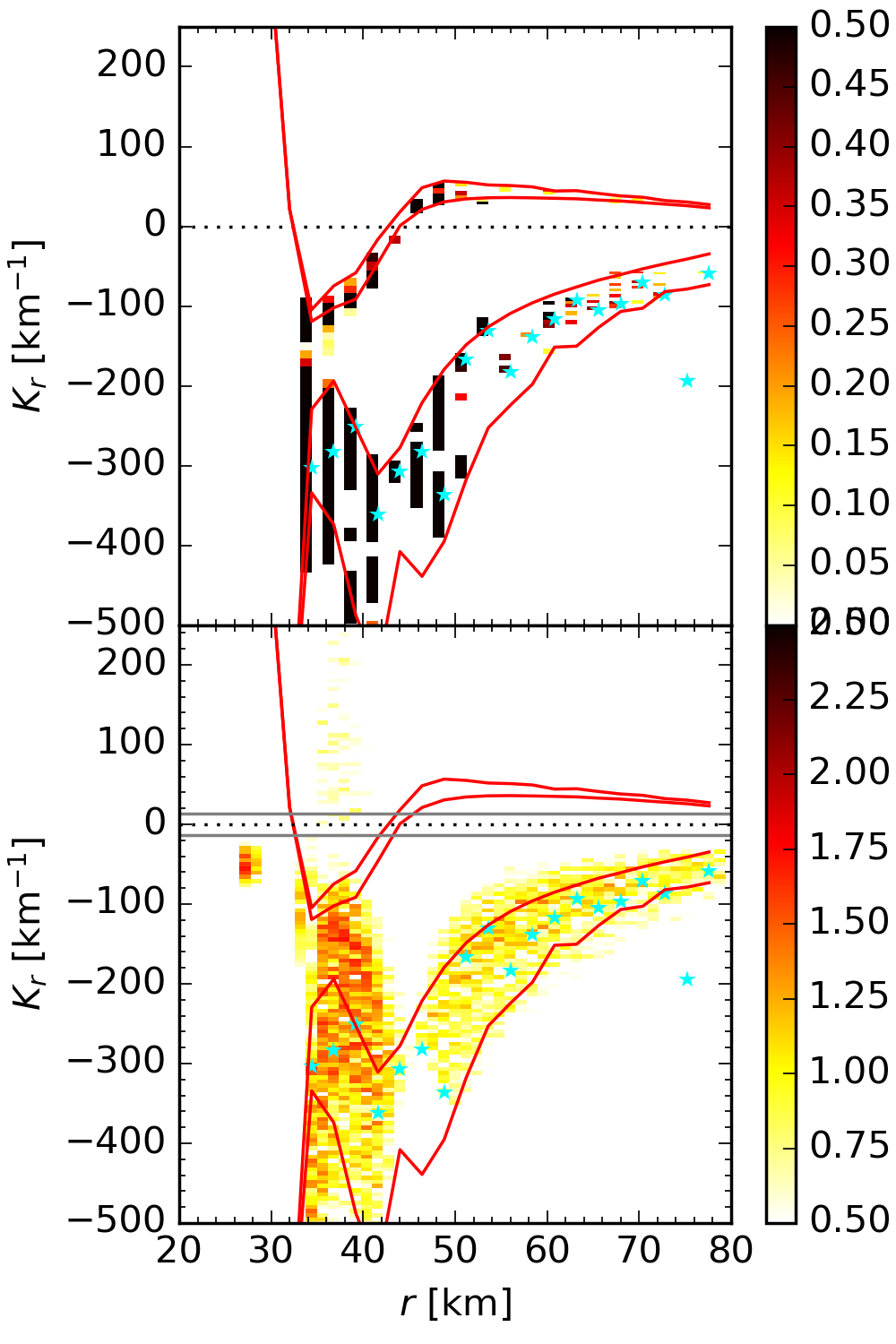}
\includegraphics[width=.24\textwidth]{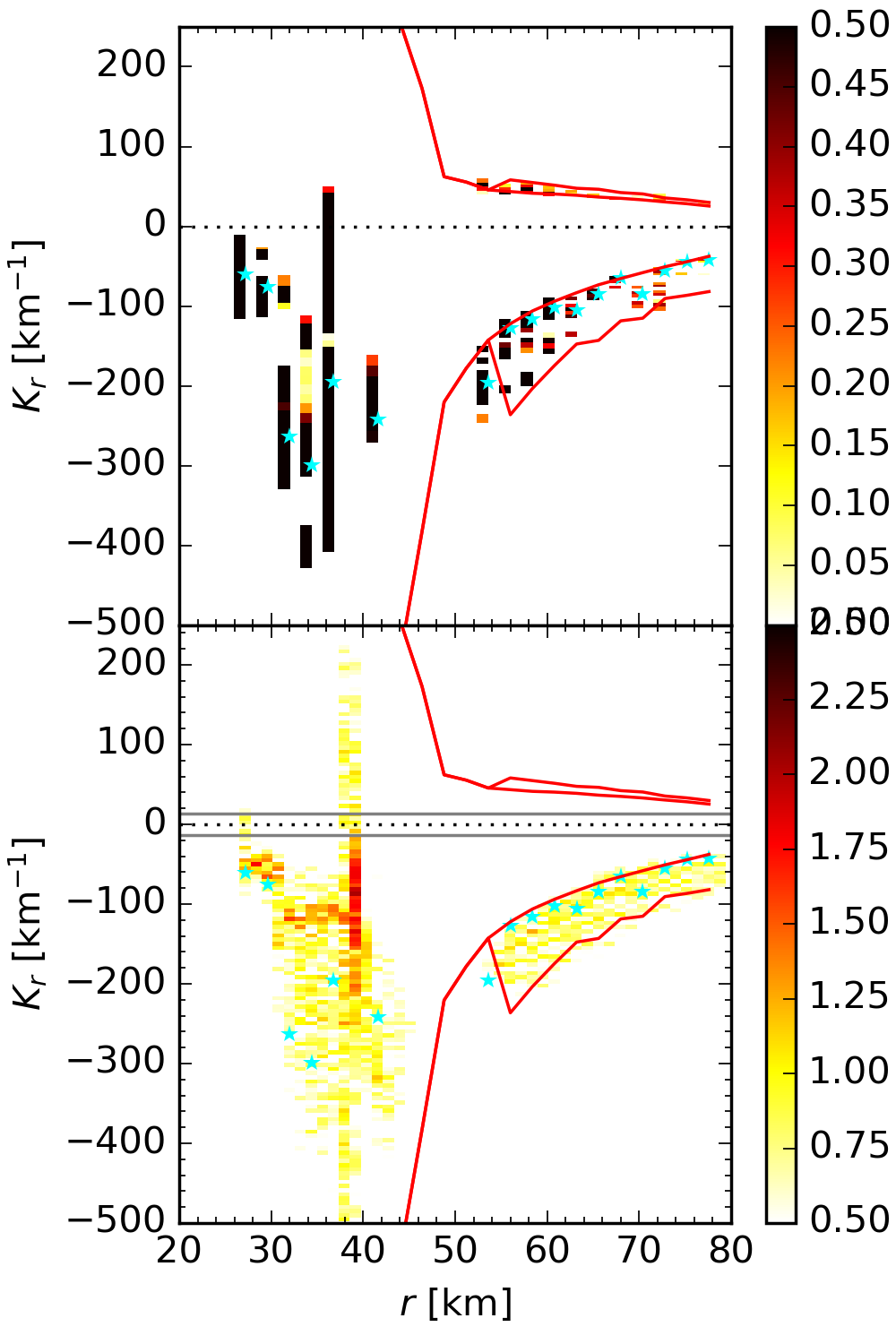}
\llap{\parbox[b]{6.1in}{\raggedright \small $b_1:$ 5\%~~~~~~~~~~~~~~~~~~~~~~~~~~~~~~~~~~~~~7\%~~~~~~~~~~~~~~~~~~~~~~~~~~~~~~~~~~~~9\%~~~~~~~~~~~~~~~~~~~~~~~~~~~~~~~~~~~~11\%\\\rule{0ex}{2.3in}}}
\caption{\label{fig:FFT_LSA} Comparison between LSA results (upper panels) and spectrograms (lower panels) at four representative times of 0.32, 0.64, 0.96, and 1.28~ms, with different $b_1$ values. The color bars are for the unstable growth rate $\mathrm{Im}(\Omega)~[{\rm km}^{-1}]$ (upper panels) and the Fourier transformed magnitude $\log_{10}\big|\mathcal F_{r,K_r} [10^{33}~{\rm cm}^{-3}]\big|$ (lower panels). 
The cyan dots show the values of $K$ with maximum growth rates calculated from LSA. The red curves correspond to the analytical edges of the unstable $K$ branches of the dispersion-relation solution (see text for details).}
\end{figure*}

Fig.~\ref{fig:FFT_LSA} compares the LSA results derived with two different methods in upper panels, and the LSA results with the spectrograms in the lower panels for four stages at 0.32, 0.64, 0.96, and 1.28~ms, respectively. 
The colored squares in the upper panels show the growth rate obtained from the discretized LSA based on the coarse-grained $\tilde G$. 
The red curves are the solutions obtained from Eq.~\eqref{eq:K_boundary}, delineating the corresponding unstable $K$ mode boundaries. 

Focus first on the spectrograms in the lower panels, where the color highlights the dominant modes of the corresponding FFC flavor wave number $K$, which are highly inhomogeneous (note that this $K$ is different from the shifted $\tilde K$ in LSA).
For the first two stages shown in Fig.~\ref{fig:FFT_LSA}, the magnitude of the dominant $K$ gradually increases from $\sim 100~{\rm km}^{-1}$ at outer radii to a few hundreds at inner radii. 
These wave numbers far exceed the maximum value that can be resolved with a limited number of radial grid points, e.g., $N_r=250$ as adopted in the ECT models, highlighting the necessity of using sufficiently fine radial grids in QKE simulations even when attenuation factors are applied.
In the latter two stages (third and fourth panels in the bottom row) when the FFC region extends deeper inside the neutrinosphere, the range of dominant $K$ modes is extended to lower $|K|$ values at $\sim 30-40$~km.
Note that near $\sim 38$~km at 1.28~ms, the dominant vertical stripe is related to the emergence of the E-XLN zero surface and the associated flavor swap. 

Turning to the LSA results, as seen in the first two panels, both LSA methods predict two branches of unstable axisymmetric modes: one above zero and the other one below.
Compared to the negative branch, the positive branch has a narrower range of $K$ and lower growth rates.
The dispersion relation results are in good agreement with the discretized LSA.
Both branches are dominated by the near-luminal modes so that $\tilde \Omega$ approximately equals $\tilde K$ \cite{yi2019dispersion,capozzi2019fast}.
Unless $\tilde K$ exactly cancels $\sqrt{2}G_F I_1$, the wave number $K$ has to be of a similar order of magnitude as the E-XLN flux term.
This once again supports the necessity of providing sufficient radial resolution.
For the latter two snapshots, both methods are still in good agreement with each other at the outer radii.
As the FFI moves further inside the neutrinosphere at the inner radii at 0.96~ms, the angular crossing point $v_c$ is far away from $v_r=1$ so that the unstable branches cover a broader range of $K$, which may interfere each other and generate new modes outside the boundaries but between the two branches.
When the total E-XLN turns negative at the inner radii at 1.28~ms, the small side becomes connected to $v_r=-1$, which is related to the uncaptured unstable modes.

Comparing LSA results to the spectrogram, the negative $K$-branch perfectly overlaps with the dominant modes in the spectrogram for all stages.
In contrast, spectrogram analysis does not result in any substantial weights in the positive-$K$ branch predicted by LSA. 
This implies that even if there exist multiple unstable branches, they may not have equal physical significance.
For the same reason, one could also speculate that if symmetry-breaking modes that are axially asymmetric exist, their physical relevance would depend on the detailed features of their branch and the associated growth rates.
If the growth rates of these symmetry-breaking modes are much smaller than those of the dominant axisymmetric modes, they will likely remain subdominant during the nonlinear development of FFC. 
Whether any significant axial asymmetry can develop requires further studies beyond the scope of this paper.
Note that even at the last stage where the $K$-ranges cannot be well predicted by Eq.~\eqref{eq:K_boundary} at inner radii, the spectrogram broadly agrees with results obtained with discretized LSA.

\section{Impact of attenuation factor}
\label{sec:attenuation}
In this section, we discuss the impact of the artificial attenuation of $H_{\nu\nu}$ on each of the three evolution episodes.

\begin{figure}
    \centering
    \includegraphics[width=0.98\columnwidth]{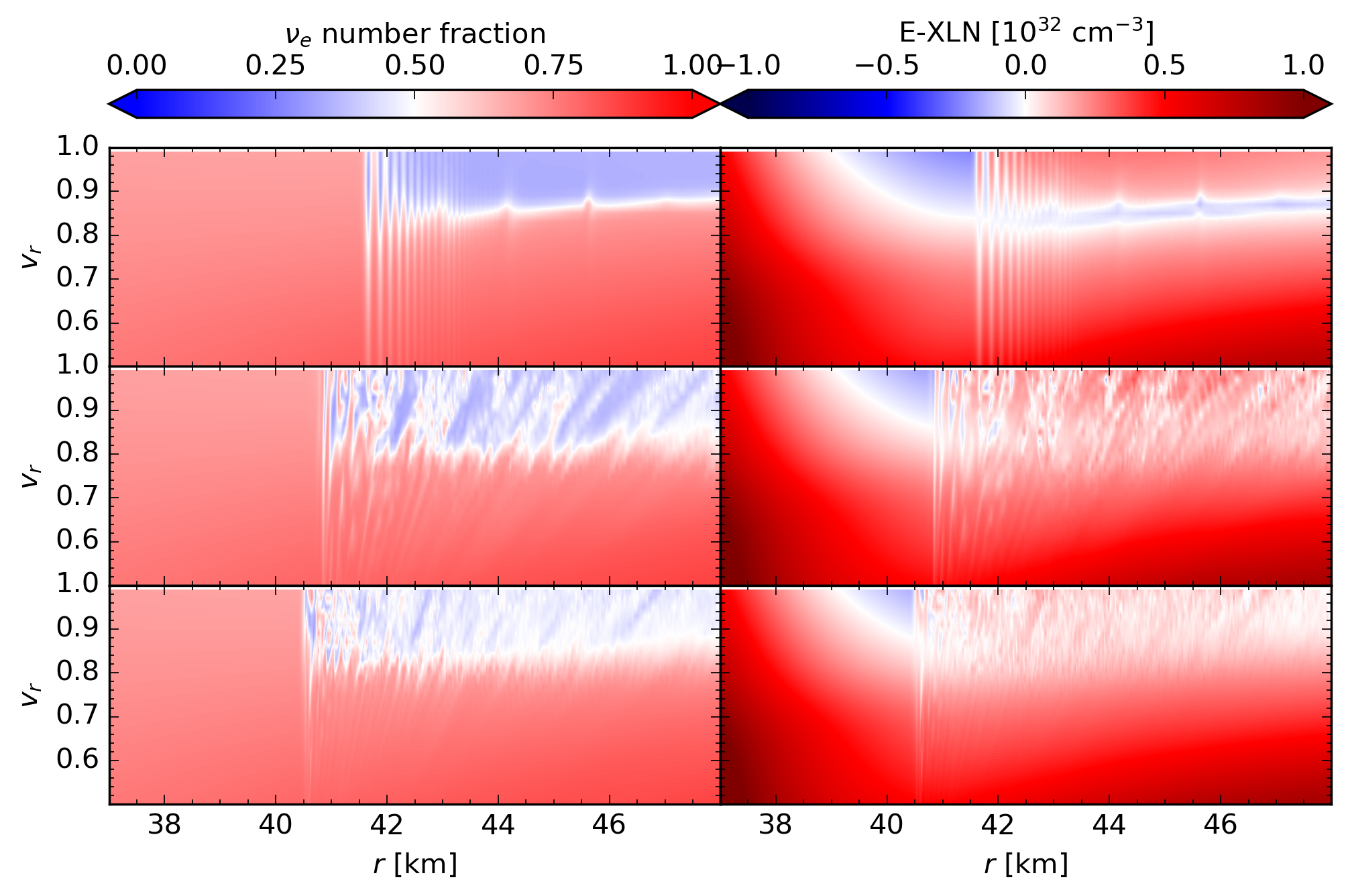}
		\llap{\parbox[b]{5.6in}{\small QKE-L\\\vspace{.4in}QKE-M\\\vspace{.4in}QKE-H\\\rule{0ex}{0.3in}}}
    \caption{Evolution of $\nu_e$ number fraction (first column) and E-XLN angular distribution (second column) near 40~km for three models including QKE-L, QKE-M, and QKE-H at $t=0.16$~ms. The $v_r$ range from 0.5 to 1 is particularly focused to show the different onset radii of flavor conversion more clearly.}
    \label{fig:details_stage_2}
\end{figure}

\begin{figure}
    \centering
    \includegraphics[width=\columnwidth]{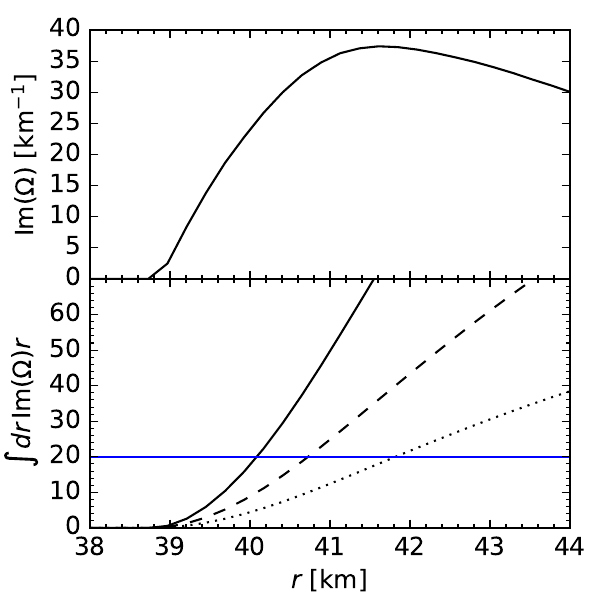}
    \caption{The maximum growth rate ${\rm Im}(\Omega)$ for Model QKE-H (upper panel) and the corresponding radial integral as a function of radius (lower panel) for Models QKE-H (solid), QKE-M (dashed), and QKE-L (dotted). The horizontal blue line in the bottom panel shows a scale close to $-\ln \theta_V$, quantifying the amplification factor required for the seed perturbation to reach the nonlinear regime.}
    \label{fig:shallow_crossing_LSA}
\end{figure}

\subsection{Shallow-crossing episode}
For the shallow-crossing episodes, the growth rate of FFCs is generally associated with the geometrical mean of $|I_>|$ and $|I_<|$ (Eq.~\eqref{eq:relaxation_tau}).
When the E-XLN first appears at the beginning of the shallow-crossing episode, $|I_<|$ is small. 
In this case, applying attenuation can artificially reduce the FFI growth rate such that the small off-diagonal elements seeded by vacuum mixing take a substantial distance to grow to the nonlinear regime.
Hence, the onset radius of FFC in attenuated QKE models may be affected by the adopted attenuation factor, which was already hinted by Fig.~\ref{fig:rp__a421}, where the FFC onset radius is smaller in Model QKE-H or Model QKE-M than that in Model QKE-L.

This effect is shown more clearly in the zoomed-in panels of Fig.~\ref{fig:details_stage_2} for better illustration.
In Model QKE-L with the most attenuation, the onset radius is about 42~km, whereas the values are $\sim 41$~km in Model QKE-M and $\sim 40$~km in Model QKE-H. 
Fig.~\ref{fig:shallow_crossing_LSA} further shows the maximal growth rate in QKE-H model in the top panel and its integration over radius from where the angular crossing first appears in the bottom panel.
Additionally, we compare in the bottom panel the integrated quantity to 20, a value close to $-\ln \theta_V$, which represents the required amplification for the flavor mixing seed triggered by vacuum oscillations to reach the nonlinear regime.
The solid curve is for Model QKE-H, whereas dashed and dotted curves are for Models QKE-M and QKE-L with stronger attenuation.
The intersection radii are consistent with the onset radii found in the simulations shown in Fig.~\ref{fig:details_stage_2}, confirming that the exact onset time and radius in the shallow-crossing episodes are slightly affected by the attenuation scheme.

In addition to the onset radius, Fig.~\ref{fig:details_stage_2} also shows that both the initial flavor overconversion and the size of the region where E-XLN crossings exist without triggering flavor conversion decrease when less attenuation is applied.
This indicates that in the case without any attenuation, the elimination of the E-XLN crossings by flavor equipartition on the small side, which is the basis of the ECT subgrid model that will be discussed in Sec.~\ref{sec:ect_result}, should be a good approximation.

\subsection{Near-crossing-elimination episode}
Recall from Sec.~\ref{subsec:elimination} that this episode is sustained by a delicate balance between the FFI growth, regulated by $|I_<|$, and the collisional rates.
Naively, one might expect that applying the attenuation while $|I_<|$ is small would disrupt this balance by suppressing the growth rate of the instability, thereby delaying or weakening the flavor conversion.
However, panels (b) and (c) of Figs.~\ref{fig:lams_iz13_nopre} and \ref{fig:lams_iz18_nopre} show that the three attenuated models give rise to very similar coarse-grained results throughout this episode, indicating that the dynamical balance is self-maintained rather than being disrupted.

\begin{figure*}
    \centering
    \includegraphics[width=0.49\linewidth]{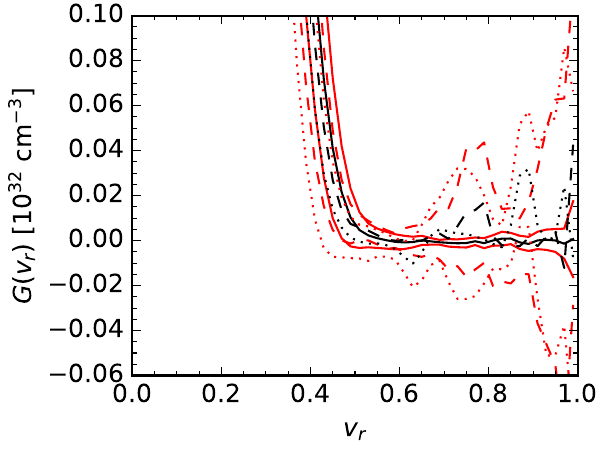}
		\llap{\parbox[b]{4.6in}{\small $t=0.96$~ms\\\rule{0ex}{2.3in}}}
    \includegraphics[width=0.49\linewidth]{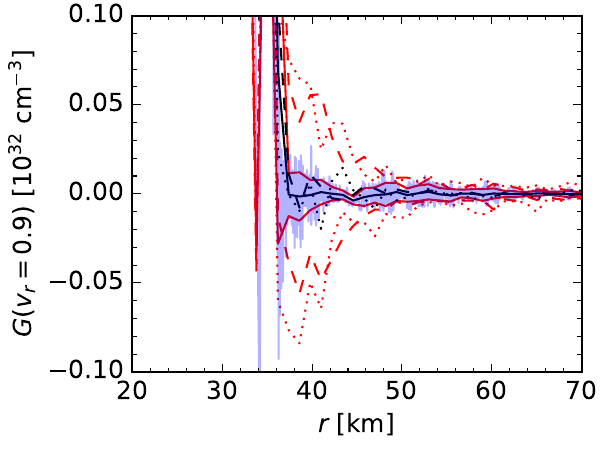}
    \includegraphics[width=0.49\linewidth]{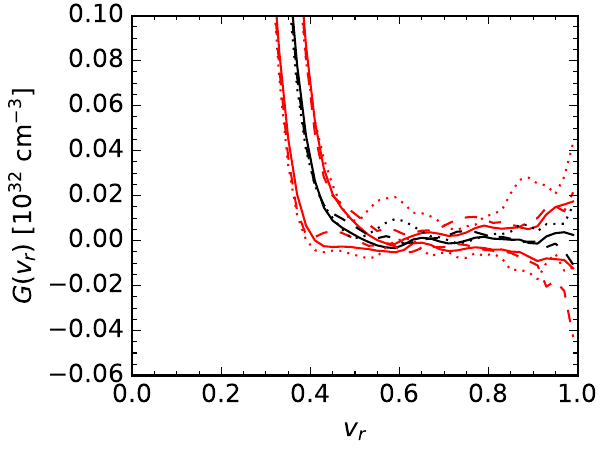}
		\llap{\parbox[b]{4.6in}{\small $t=1.28$~ms\\\rule{0ex}{2.3in}}}
    \includegraphics[width=0.49\linewidth]{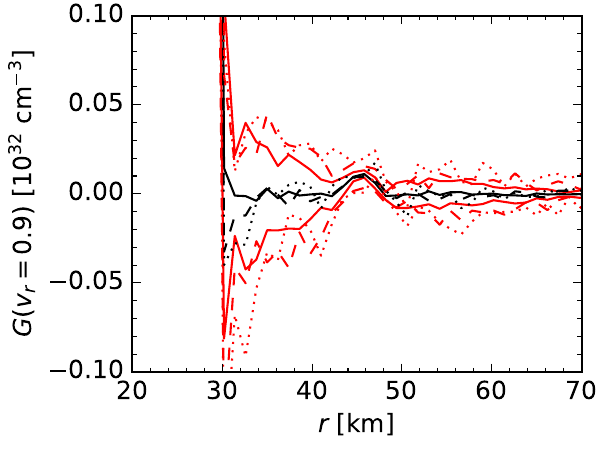}
    \caption{Coarse-grained E-XLN angular distribution $\tilde G(v_r)$ (black) and standard deviation (red) of $G(v_r)$ at $r=41$~km (left panels) and their values at $v_r=0.9$ as a function of radius (right panels) for models QKE-L (dotted), QKE-M (dashed), and QKE-H (solid) at t=0.96 (upper panels) and 1.28~ms (lower panels). 
    The original profiles of all $G$ for Model QKE-H are shown by blue curves in the upper right panel.}
    \label{fig:G_variation}
\end{figure*}

Fig.~\ref{fig:G_variation} shows in detail how this balance is maintained by comparing models with different levels of attenuation. 
The left panels show the coarse-grained $\tilde G(v_r)$ (black curves), along with the standard deviation computed based on all  $G(v_r)$ inside the subgrid with width $\Delta r$ (red curves) at 41~km for $t=0.96$ (upper panel) and 1.28~ms (lower panel).
Comparing the solid (QKE-H), dashed (QKE-M), and dotted (QKE-L) curves clearly shows that the variation of $G$ on the small side increases with stronger attenuation on $H_{\nu\nu}$. 
While it is not obvious from these two panels, Fig.~\ref{fig:lams_iz18_nopre} also shows that $|I_<|$ generally reduces with less attenuation. 

The right panels of Fig.~\ref{fig:G_variation} further show $\tilde G(v_r=0.9)$ and the associated standard deviation as functions of radius, confirming the same trend.
This behavior reflects the balance condition of marginal FFCs: When the attenuation is stronger, a larger value of $|I_<|$ is required to maintain the effective growth rate that balances the collisional rate, and the angular structure on the small side correspondingly becomes more fluctuating.

\subsection{Swapping episode}
We also inspect how the attenuation factor applied to $H_{\nu\nu}$ impacts the propagation of the E-XLN zero-surface front in the swapping episodes.
The left panels of Fig.~\ref{fig:rp__a421} show that the swapping interface propagates earlier or faster in Model QKE-L than in Model QKE-H.
Similarly, Figs.~\ref{fig:lams_iz18_nopre} and \ref{fig:lams_iz25_nopre} indicate that, in Model QKE-H, the fronts reach $41~\text{km}$ and $50~\text{km}$ later than in the other two models.
This delayed propagation is likely related to the later onset of the transition into negative E-XLN at inner radii in Model QKE-H.
As shown in Fig.~\ref{fig:lams_iz13_nopre}, the transition to negative E-XLN occurs slightly earlier in Model QKE-L than in Model QKE-H.
This earlier transition in Model QKE-L naturally facilitates an earlier onset of the zero-surface front.

We speculate that this trend is related to the self-regulated balance discussed in the previous subsection.
Under weaker attenuation, the relatively more efficient flavor conversion pins the system more tightly to the crossing-eliminated, flavor-equilibrated configuration, holding the coarse-grained E-XLN close to zero with only small residual fluctuations.
Stronger attenuation instead reduces the ratio between the flavor conversion and the collisional rates, leading to larger residual deviations and fluctuations of the E-XLN angular distribution (see Fig.~\ref{fig:G_variation}).
Such fluctuations allow local domains with negative E-XLN to emerge more readily, seeding the zero surface and thereby triggering the swapping earlier.
In this sense, the earlier onset of the swapping in the more attenuated models is an artifact of the attenuation, and the onset is expected to occur even later in the unattenuated limit.

\section{Comparison with two-step model and ECT solutions}
\label{sec:comparison}
In this section, we first compare in Sec.~\ref{subsec:quasistationary} the time-dependent QKE evolution with the quasistationary solutions obtained from two-step simulations in each stage. 
Then, Sec.~\ref{sec:ect_result} compares time-dependent QKE solution with the subgrid ECT models, and Sec.~\ref{subsec:robustness} examines the robustness of the ECT results against the choices of the prescription form and the relaxation timescale.

\subsection{Quasistationary states of separate stages}
\label{subsec:quasistationary}
Besides the staged QKE evolution discussed above, we also perform additional QKE simulations for each stage \emph{separately} under static background profiles given by each stage. 
For each of them, the initial condition is determined by the asymptotic state of pure classical transport without considering flavor oscillations under the same static background as done in Ref.~\cite{xiong2025occurrence}, which generates deep E-XLN crossings similar to those indicated by the star symbols.
For each of them, we evolve the QKE using the same attenuation factor corresponding to the QKE-H model until the system relaxes to its quasistationary state.
The final values of $I_>$ and the number densities of different species are shown by hollow circles in Figs.~\ref{fig:lams_iz13_nopre}, \ref{fig:lams_iz18_nopre}, and \ref{fig:lams_iz25_nopre} for nine different stages. 
Comparing them to the time-evolving QKE solution (solid curve) shows general agreement.
Substantial discrepancy only occurs at, e.g., $t=2.56$~ms for $r=36$~km and $t=1.9$~ms for $r=41$~km.

\begin{figure*}[!hbt]
\centering
\includegraphics[width=0.98\textwidth]{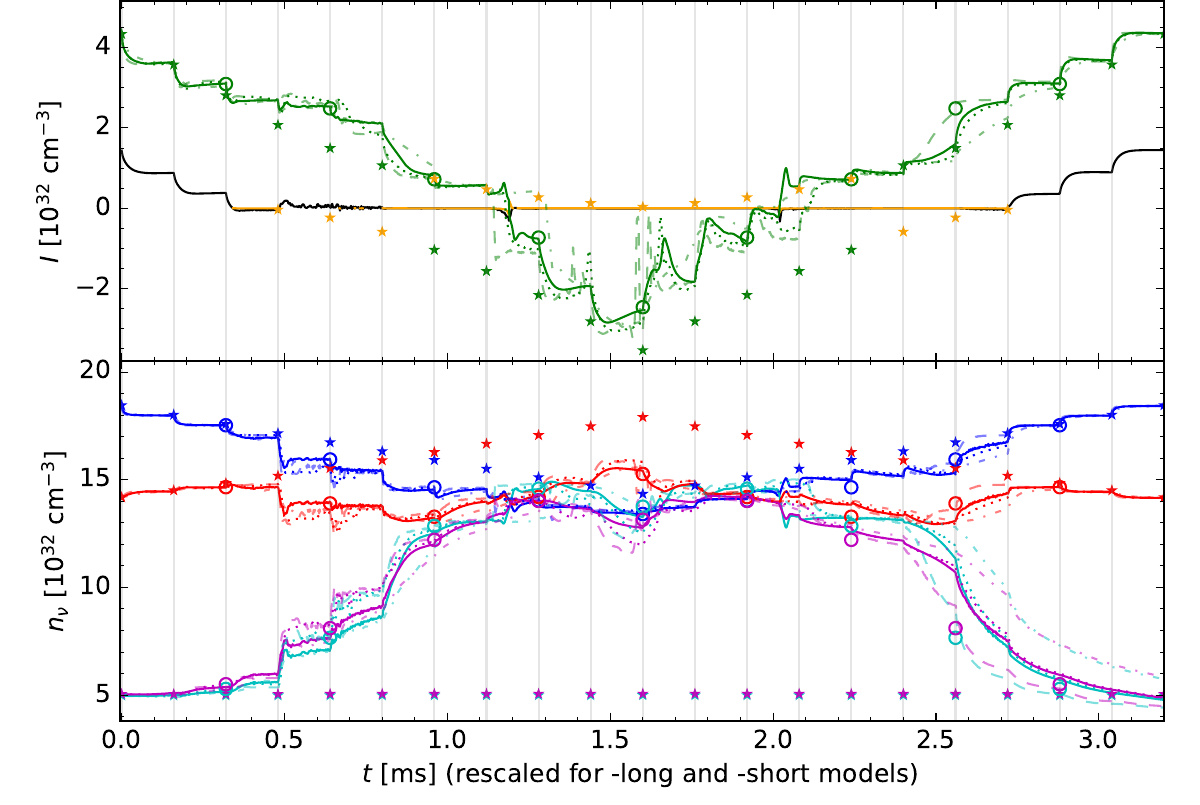}
\llap{\parbox[b]{6.2in}{\raggedright \small $b_1:$\\\rule{0ex}{4.45in}}}
\llap{\parbox[b]{6.25in}{\raggedright \small 4\%~~~5\%~~~6\%~~~7\%~~~8\%~~~9\%~~10\%~11\%~~12\%~~13\%~14\%~~13\%~~12\%~11\%~~10\%~~~9\%~~~8\%~~~7\%~~~6\%~~~5\%\\\rule{0ex}{4.3in}}}
\llap{\parbox[b]{13.2in}{\small (a)\\\vspace{2.05in}(b)\\\rule{0ex}{2.3in}}}
\caption{\label{fig:lams_iz13_nopre_long_short} Same as Fig.~\ref{fig:lams_iz13_nopre} except that Models ``QKE-L-long'' (transparent dashed) and ``QKE-H-short'' (transparent dash-dotted) are compared. The horizontal time axis is rescaled so that their transitions align with Models ``QKE-H'' (solid) and ``QKE-L'' (dotted).}
\end{figure*}

To pin down the underlying reason that causes the discrepancy, we perform two more sets of time-dependent QKE simulations that adopt different stage durations: one with $t_{\rm stage}=0.32$~ms for the QKE-L set-up and the other one with $t_{\rm stage}=0.08$~ms for the QKE-H set-up. 
These two auxiliary models are named ``QKE-L-long'' and ``QKE-H-short'', respectively, and are listed in Table~\ref{tab:parameters}.
In Fig.~\ref{fig:lams_iz13_nopre_long_short}, we compare results of the QKE-H-short (dash-dotted curves) and QKE-L-long (dashed curves) models with those of the QKE-H model (solid curves) as well as of the two-step models (hollow circles) taken at 36~km.
Clearly, model QKE-H-short deviates significantly more from the fixed-background simulations, particularly during the reversal half phase. 
In contrast, Model ``QKE-L-long'' shows much better agreement with the two-step quasistationary results.  

This comparison shows that the system requires a finite relaxation time to reach the quasi-steady state of the two-step model with fixed background. 
When $t_{\rm stage}$ is too short, the system is forced to evolve before reaching this configuration.
For example, at $t = 2.56~\text{ms}$ mentioned above, taking a longer $t_{\rm stage}$ in the time-dependent QKE model allows sufficient time for the heavy-lepton flavor neutrinos that are produced by FFCs in earlier stages to diffuse and escape from the neutrino spheres. 
In this case, the system has enough time to evolve toward the corresponding quasi-steady state of the two-step model, resulting in better agreement of the results.

\subsection{Effective classical transport results}
\label{sec:ect_result}
\begin{figure*}[!hbt]
\centering
\includegraphics[width=0.98\textwidth]{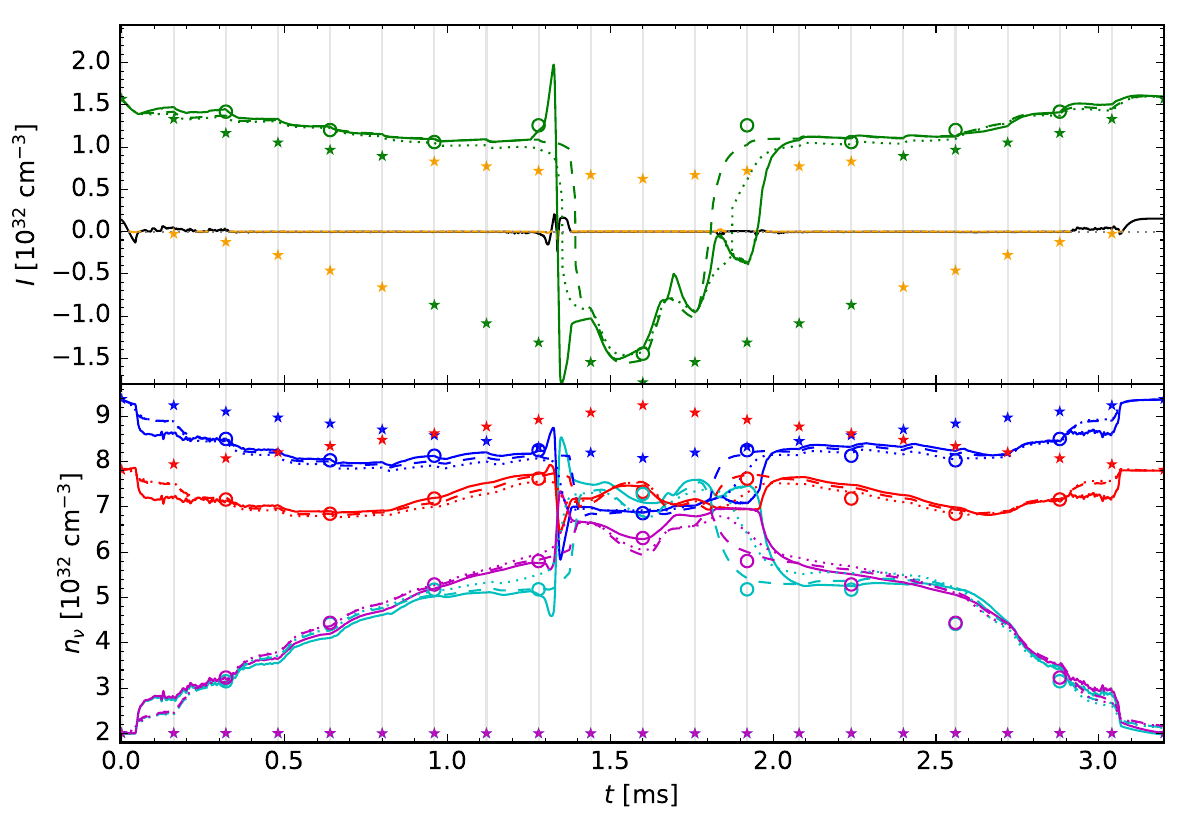}
\llap{\parbox[b]{6.2in}{\raggedright \small $b_1:$\\\rule{0ex}{4.45in}}}
\llap{\parbox[b]{6.25in}{\raggedright \small 4\%~~~5\%~~~6\%~~~7\%~~~8\%~~~9\%~~10\%~11\%~~12\%~~13\%~14\%~~13\%~~12\%~11\%~~10\%~~~9\%~~~8\%~~~7\%~~~6\%~~~5\%\\\rule{0ex}{4.3in}}}
\llap{\parbox[b]{13.2in}{\small (a)\\\vspace{1.95in}(b)\\\rule{0ex}{2.4in}}}
\caption{\label{fig:lams_pre} Same as Fig.~\ref{fig:lams_iz18_nopre} except that solid, dashed, and dotted curves are for the ``QKE-H'', ``ECT'', and ``ECT-a0.1'' simulations, respectively. All quantities are evaluated at 41~km.}
\end{figure*}

Fig.~\ref{fig:lams_pre} compares  the evolution of quantities at 41~km from the QKE-H model (solid curves) to results obtained in the ECT (dashed curves) and ECT-a0.1 (dotted curves) models, which use the power-1/2 prescription (Eq.~\eqref{eq:p_sur}) with different relaxation time scales (Eqs~\eqref{eq:relaxation_tau} and \eqref{eq:relaxation_tau1}).  
The evolutions in ECT models agree well with the QKE-H results particularly during the near-crossing-elimination episode. 
This suggests that under the dynamical balance discussed in Sec.~\ref{subsec:elimination}, the length scale associated with the growth of the flavor instability $l_{\rm osc}$, defined as the inverse of the maximum growth rate ${\rm Im}(\Omega)$ over all $K$ modes obtained from the LSA in Sec.~\ref{sec:spectrogram}, is generally smaller than $l_{\rm env}$, which is the shortest spatial scale height of the environment, such that the implicit assumption of $l_{\rm osc}\lesssim l_{\rm env}$ underlying the ECT subgrid framework is fulfilled. 
The same agreement holds over the entire radial range: the right panels of Fig.~\ref{fig:rp__a421} (introduced in Sec.~\ref{subsec:radial}) compare the number density ratios $R_{\nu_i}(r,t)$ obtained in the ECT models to the QKE-H results, showing remarkable quantitative agreement for $0.32$~ms$\leq t\leq 1.12$~ms, particularly during the near-crossing-elimination episode.

During the shallow-crossing episodes, slightly larger differences between the ECT models and the QKE-H model appear in the very beginning and close to the end of the evolution.  
This suggests that $l_{\rm osc}\gtrsim l_{\rm env}$ in these phases in the QKE-H model, so that the local conditions have evolved before the flavor waves are developed. 
However, comparing Fig.~\ref{fig:lams_pre} with Fig.~\ref{fig:lams_iz18_nopre} reveals that these differences between the ECT and QKE models are smaller when less attenuation is applied, consistent with what was reported in Ref.~\cite{xiong2025robust}.
This implies that these small differences may actually originate from the attenuation, and could be even smaller if no attenuation is applied at all.

The differences between the QKE models and the ECT models are substantially larger during the swapping episodes.
The right panels of Fig.~\ref{fig:rp__a421} show that the formation and propagation of the sharp E-XLN zero surface obtained in QKE-H model cannot be well reproduced by the ECT models. 
This disagreement in fact has a more fundamental origin: the dynamics of the swapping episodes is not driven by the E-XLN angular crossings, on which the subgrid prescriptions are exclusively based, but by the coherent flavor-swap mechanism across the E-XLN zero surface~\cite{zaizen2024fast} discussed in Sec.~\ref{subsec:swapping}.
There, the nearly complete flavor conversion is facilitated by the large off-diagonal coherence $\langle\varrho_{ex}\rangle_E$ built up over the $v_r$-independent zero surface (Fig.~\ref{fig:swapping_2d}), a genuinely coherent phenomenon with no counterpart in the ECT framework, in which only the diagonal elements are evolved and the flavor content is redistributed only in response to E-XLN crossings.
Consequently, although the ECT models do develop sharp transitions in various quantities that superficially resemble a swapping interface [see Figs.~\ref{fig:rp__a421} and \ref{fig:lams_pre}], these arise from the crossing-based redistribution acting on the rapidly changing profiles rather than from a true flavor swap: no flavor-swap signature of the kind displayed in Fig.~\ref{fig:swapping_2d} is present in the ECT runs at all.

\subsection{Robustness of the ECT model}
\label{subsec:robustness}
We now turn to an important question, namely how robust the ECT results are with respect to the specific choices made within the subgrid framework.
Our three ECT models test the two main ingredients separately: the functional form of the prescription (the power-1/2 form in Models ECT and ECT-a0.1 versus the box form in Model ECT-box), and the relaxation timescale [Eq.~\eqref{eq:relaxation_tau} in Models ECT and ECT-box versus the $10^4$ times longer $\tau'$ of Eq.~\eqref{eq:relaxation_tau1} in Model ECT-a0.1].
The right panels of Fig.~\ref{fig:rp__a421} show that the three models yield nearly identical radial profiles of $R_{\nu_i}$ throughout most of the evolution, demonstrating that the coarse-grained outcome is largely insensitive to both choices.

The insensitivity to the functional form, previously found for stationary backgrounds in Ref.~\cite{xiong2025robust}, can be understood as follows.
In the near-crossing-elimination episode, the small side of the E-XLN distribution corresponds to $|I_<|$ being very close to zero.
Under such conditions, both types of prescription naturally yield a survival probability close to unity on the larger side of the distribution.
As a result, the dominant flavor component remains largely preserved, while the small side is efficiently redistributed.
The E-XLN is therefore carried outward and redistributed among neutrinos propagating along different angular directions in a differential manner.
This mechanism does not strongly depend on the detailed functional form assumed in the prescription.

Subtle differences nevertheless remain.
Comparing Models ECT-box and ECT, small deviations appear mainly at larger radii outside the neutrinosphere, when the angular distribution is more forward peaked, and the outcome is more sensitive to the prescription near the crossing angle, which is again consistent with the results reported in \cite{xiong2025robust}. 
The differences between Models ECT-a0.1 and ECT are even smaller during the shallow-crossing and near-crossing-elimination episodes. 
While the much longer relaxation timescale shifts the timing and sharpness of the swapping interface, the flavor swapping phenomenon cannot be accurately captured by any ECT models in any case as discussed above.

These findings indicate that accurately capturing the existence and evolution of the angular crossing is perhaps more crucial than modeling the fine details of the angular distribution or the precise relaxation timescale within the prescription framework.
As long as the prescription correctly identifies the presence of the crossing and regulates $|I_<|$ toward marginal stability, it can reproduce the main features of the flavor evolution.
The precise shape of the assumed angular transition plays a secondary role compared to the correct treatment of the instability condition itself.
Hence, while the validity of implementing instantaneous flavor redistribution in hydrodynamic simulations is supported, resolving the angular crossing is required.

\section{Conclusions}
\label{sec:conclusion}
To explore how FFCs develop and evolve in a time-dependent environment, we have extended our previous study based on a stationary, spherically symmetric supernova background by introducing staged evolution of its $Y_e$ profile. 
This framework allows FFCs to evolve continuously in response to the changing environment. 
In contrast to the widely used two-step approach, no flavor instability is imposed in the initial condition: The E-XLN crossings arise and develop marginally through the self-consistent interplay of collisions, transport, and flavor conversion.
We have implemented three different $H_{\nu\nu}$ attenuation levels, and performed linear stability and spectrogram analyses to characterize the physical scales associated with the flavor waves.

The overall evolution can be broadly classified into three episodes: the shallow-crossing episode, the near-crossing-elimination episode, and the swapping episode.
During the shallow-crossing episode, once the angular crossing is established, FFCs begin to operate efficiently.
The characteristic scales of the flavor waves identified by the spectrogram are highly consistent with the negative unstable branch predicted by the LSA.
Interestingly, the other unstable branch predicted by LSA does not appear in the spectrogram of the simulations, indicating that not all linearly unstable modes necessarily leave a dynamical imprint.

In the near-crossing-elimination episode, the balance between collisions and FFCs keeps the system in a near-quasistationary state in which the emerging crossings are continuously eliminated, so that the strongly unstable regime adopted in two-step models is never reached while substantial flavor conversion is nevertheless achieved. 
The oscillation length scale becomes progressively smaller in this episode, and the spectrogram and LSA remain in good agreement in this regime. 
We find that collisional decoherence plays a key role in driving the system toward crossing elimination by suppressing off-diagonal flavor coherence while simultaneously flattening the small side of the angular distribution.

In the swapping episodes, where the E-XLN reverses sign, flavor swap develops across a dynamically propagating E-XLN zero surface.
A distinctive feature identified in this work is that the E-XLN zero surface is not fixed but evolves dynamically, due to the use of realistic and self-consistent collisional rates as the neutrino source in the setup. 
Moreover, comparing the swapping episodes in the two halves of the background evolution reveals a clear breaking of the time-reversal symmetry: In the reverse phase, the zero surface is less pronounced and propagates inward instead, highlighting the intrinsically out-of-equilibrium nature of the system.

We have also explored the impact of the adopted attenuation factors on the results. 
The general evolution is largely insensitive to the choice of the attenuation factor adopted in this work.
Stronger attenuation shifts the FFC onset radius outward at the beginning of the shallow-crossing episode due to the suppressed instability growth rate. 
In the crossing-elimination stages, the dynamical balance keeping the system in the quasistationary state is maintained, although the size of the crossing depth ($|I_<|$) becomes larger with stronger attenuation.
For the swapping episodes, a larger attenuation factor leads to earlier launch of the E-XLN zero surface.
More importantly, the E-XLN crossing elimination and flavor equipartition on the small side of the crossing during the shallow-crossing and near-crossing-elimination stages hold better with weaker attenuation.

Comparing with the two-step approach, we find that the flavor content at the end of each stage generally agrees with the quasistationary solutions obtained from separate two-step simulations under the corresponding static backgrounds.
The degree of agreement depends on the pace of the background evolution: a longer stage duration allows the system to relax closer to the quasistationary states, while a shorter duration leads to larger deviations, particularly during the reversal half of the evolution.

We further examined the validity and robustness of the subgrid ECT framework.
The ECT results converge toward the QKE evolution during the shallow-crossing and near-crossing-elimination episodes, and are remarkably insensitive to both the functional form of the prescription and the adopted relaxation timescale, since the coarse-grained outcome is mainly determined by the condition of crossing elimination rather than by these details.
The exception is the swapping episodes, in which the dynamics is no longer driven by the E-XLN crossings but by the coherent flavor swap across the zero surface, a mechanism absent from the crossing-based prescriptions by construction.
This comparison suggests that a self-consistent coupling between subgrid prescriptions and classical transport can reproduce the key dynamical features of time-dependent FFCs, supporting their application in global simulations.

We conclude by noting the scope of the present study.
The current models adopt spherical symmetry and do not yet include feedback of flavor conversion onto the background matter.
We also do not observe a significant impact from the inhomogeneous mode of slow flavor instabilities, although their potential implications have been suggested recently \cite{fiorillo2025theory,fiorillo2025lepton,padilla2025flavor,fiorillo2025when}.
Signatures consistent with collisional flavor instability were observed during certain stages; their systematic investigation and interplay with FFCs are deferred to future work.
Nevertheless, the framework used here offers a clear and controlled way to identify the main mechanisms that govern time-dependent FFCs, and future extensions incorporating multidimensional effects, turbulence, and fully self-consistent neutrino--matter coupling will build naturally on this foundation.

\begin{acknowledgments}
ZX acknowledges support of the European Research Council (ERC) under the European Union’s Horizon 2020 research and innovation program (ERC Advanced Grant KILONOVA No. 885281) and under the ERC Grant (NeuTrAE, No. 101165138).
MRW acknowledges support from the National Science and Technology Council, Taiwan under Grant No.~111-2628-M-001-003-MY4, No.~115-2112-M-001-054-MY5, the Academia Sinica under Project No.~AS-CDA-109-M11, 
and the Physics Division of the National Center for Theoretical Sciences, Taiwan.
This research was supported in part by the cluster computing resource provided by the IT Department at the GSI Helmholtzzentrum f\"ur Schwerionenforschung, Darmstadt, Germany, and by the Gauss Center for Supercomputing (GCS) for using the GCS Supercomputer JUWELS at the Jülich Supercomputing Center, Jülich, Germany.
We acknowledge the following software: \textsc{matplotlib}~\cite{matplotlib}, \textsc{numpy}~\cite{numpy}, and \textsc{scipy}~\cite{scipy}.

The work is partially funded by the European Union. Views and opinions expressed are however those of the author(s) only and do not necessarily reflect those of the European Union or the European Research Council Executive Agency. Neither the European Union nor the granting authority can be held responsible for them.
\end{acknowledgments}

\bibliographystyle{apsrev4-1}
\bibliography{main.bbl}

\appendix

\section{Time evolution of neutrino kinetics at 50~km}
\label{sec:A_time_evolution}
\begin{figure*}[!hbt]
\centering
\includegraphics[width=0.98\textwidth]{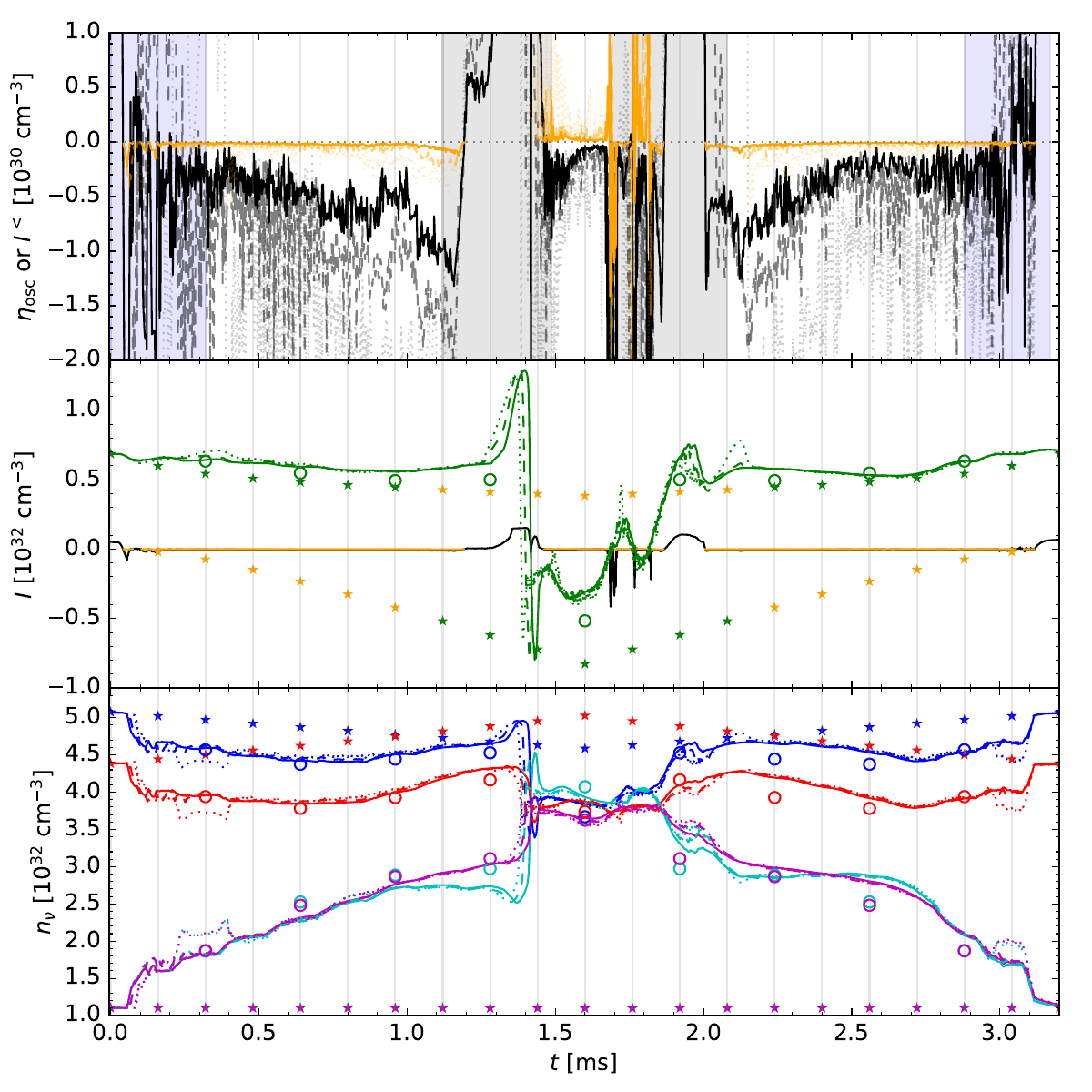}
\llap{\parbox[b]{6.2in}{\raggedright \small $b_1:$\\\rule{0ex}{4.45in}}}
\llap{\parbox[b]{6.25in}{\raggedright \small 4\%~~~5\%~~~6\%~~~7\%~~~8\%~~~9\%~~10\%~11\%~~12\%~~13\%~14\%~~13\%~~12\%~11\%~~10\%~~~9\%~~~8\%~~~7\%~~~6\%~~~5\%\\\rule{0ex}{4.3in}}}
\llap{\parbox[b]{13.4in}{\small (a)\\\vspace{1.9in}(b)\\\vspace{1.95in}(c)\\\rule{0ex}{2.35in}}}
\caption{\label{fig:lams_iz25_nopre} Same as Fig.~\ref{fig:lams_iz13_nopre} except that all quantities are evaluated at 50~km.}
\end{figure*}

Fig.~\ref{fig:lams_iz25_nopre} shows the time evolution of the same characteristic quantities as in Figs.~\ref{fig:lams_iz13_nopre} and \ref{fig:lams_iz18_nopre}, evaluated at $r=50$~km, outside the $\nu_e$ sphere.
Since FFCs first develop around and beyond the $\nu_e$ sphere, this radius is affected by flavor conversion almost immediately after the simulation starts, as indicated by the rapid initial drop of $n_{\nu_e}$ and $n_{\bar\nu_e}$ and the rise of the heavy-lepton species in panel (c).
Throughout most of the evolution, $I_<$ remains close to zero while $\eta_{\rm osc}$ stays small and negative with visible fluctuations [panel (a)], showing that the near-crossing-elimination behavior dominates at this radius.
In contrast, without oscillations, the E-XLN crossing would continuously deepen and the overall E-XLN would flip sign during the most neutron-rich stages, as shown by the orange and green star symbols in panel (b).
When FFCs operate, the sign flips of the E-XLN are delayed and appear as large transient excursions of $I_>$ at $t\approx 1.4$~ms and near $t\approx 1.9$--$2.0$~ms, whose distinct shapes reflect the time-reversal asymmetry discussed in Sec.~\ref{subsec:swapping}.
Between the two swapping episodes ($t\approx 1.4$--$1.9$~ms), FFCs drive the number densities of all four species to comparable values [panel (c)], in contrast to the no-oscillation case in which $n_{\nu_x}$ remains unchanged (purple stars).
The hollow circles obtained from the per-stage simulations generally track the continuous evolution, with the largest deviations occurring around the swapping episodes.

\end{document}